\documentclass[fleqn,usenatbib]{mnras}

\usepackage{newtxtext,newtxmath}
\usepackage[T1]{fontenc}

\DeclareRobustCommand{\VAN}[3]{#2}
\let\VANthebibliography\thebibliography
\def\thebibliography{\DeclareRobustCommand{\VAN}[3]{##3}\VANthebibliography}

\usepackage{graphicx}	
\usepackage{amsmath}	
\usepackage{amssymb}	

\usepackage{ulem}
\usepackage{xcolor}

\usepackage{subfigure}
\usepackage{multirow}
\usepackage{amssymb}
\usepackage{amsmath}
\usepackage{caption}

\newcommand{\kms}{\textrm{km~s$^{-1}$}}

\newcommand{\lsolar}{L$_{\odot}$}
\newcommand{\msolar}{M$_{\odot}$}
\newcommand{\ml}{M$_{\odot}$ yr$^{-1}$}
\newcommand{\mdot}{\dot{M}}

\title[Demise of SN 2005ip]{The Slow Demise of the Long-Lived SN 2005ip}
\author[O. D. Fox et al.]{Ori D. Fox$^{1,2}$, Claes Fransson$^{3}$, Nathan Smith$^{4}$, Jennifer Andrews$^{4}$, \newauthor K. Azalee Bostroem$^{5}$, Thomas G. Brink$^{6}$, S. Bradley Cenko$^{7,8}$, Geoffrey C. Clayton$^{9}$, \newauthor Alexei V. Filippenko$^{6,10}$, Wen-fai Fong$^{11}$, Joseph S. Gallagher$^{12}$, \newauthor Patrick L. Kelly$^{13}$, Charles D. Kilpatrick$^{14}$, Jon C. Mauerhan$^{6,15}$, \newauthor Adam M. Miller$^{16,17,18}$, Edward Montiel$^{19}$,  Maximilian D. Stritzinger$^{20}$, \newauthor Tam\'as Szalai$^{21,22}$, Schuyler D. Van Dyk$^{23}$\\
$^{1}$Space Telescope Science Institute, 3700 San Martin Drive, Baltimore, MD 21218, USA.\\
$^{2}$ofox@stsci.edu.\\
$^{3}$Department of Astronomy, Oskar Klein Centre, Stockholm University, AlbaNova, SE--106~91 Stockholm, Sweden.\\
$^{4}$Steward Observatory, 933 N. Cherry Ave., Tucson, AZ 85721, USA.\\
$^{5}$Department of Physics, University of California, Davis, CA 95616, USA.\\
$^{6}$Department of Astronomy, University of California, Berkeley, CA 94720-3411, USA.\\
$^{7}$NASA Goddard Space Flight Center, 8800 Greenbelt Road, Greenbelt, DMD 20771, USA.\\
$^{8}$Joint Space-Science Institute, University of Maryland, College Park, MD 20742, USA.\\
$^{9}$Dept.\ of Physics \& Astronomy, Louisiana State University, Baton Rouge, LA 70803, USA.\\
$^{10}$Miller Senior Fellow, Miller Institute for Basic Research in Science, University of California, Berkeley, CA 94720, USA.\\ 
$^{11}$Center for Interdisciplinary Exploration and Research in Astrophysics (CIERA) and Department of Physics and Astronomy, Northwestern University, Evanston, IL 60208, USA.\\
$^{12}$University of Cincinnati Blue Ash College, 9555 Plainfield Rd., Blue Ash, OH 45236, USA.\\
$^{13}$Minnesota Institute for Astrophysics, University of Minnesota, 115 Union St. SE, Minneapolis, MN 55455, USA.\\
$^{14}$Department of Astronomy and Astrophysics, University of California, Santa Cruz, CA 95064, USA.\\
$^{15}$ The Aerospace Corporation, 2310 E. El Segundo Blvd., El Segundo, CA 90245, USA.\\
$^{16}$Jet Propulsion Laboratory, 4800 Oak Grove Drive, MS 169-506, Pasadena, CA 91109, USA.\\
$^{17}$California Institute of Technology, Pasadena, CA 91125, USA.\\
$^{18}$Hubble Fellow.\\
$^{19}$SOFIA-USRA, NASA Ames Research Center, Mail Stop N232-12, Moffett Field, CA 94035-1000, USA.\\
$^{20}$Department of Physics and Astronomy, Aarhus University, Ny Munkegade 120, DK-8000 Aarhus C, Denmark\\
$^{21}$Department of Optics and Quantum Electronics, University of Szeged, H-6720 Szeged, D\'om t\'er 9., Hungary.\\
$^{22}$Konkoly Observatory, Research Centre for Astronomy and Earth Sciences, H-1121 Budapest, Konkoly Thege Mikl\'os \'ut 15-17,
Hungary.\\
$^{23}$IPAC/Caltech, Mailcode 100-22, Pasadena, CA 91125, USA.}
\begin{document}

\maketitle
\begin{abstract}

The Type IIn supernova (SN) 2005ip is one of the most well-studied and long-lasting examples of a SN interacting with its circumstellar environment.  The optical light curve plateaued at a nearly constant level for more than five years, suggesting ongoing shock interaction with an extended and clumpy circumstellar medium (CSM). Here we present continued observations of the SN from $\sim 1000-5000$~days post-explosion at all wavelengths, including X-ray, ultraviolet, near-infrared, and mid-infrared.  The UV spectra probe the pre-explosion mass loss and show evidence for CNO processing. From the bolometric light curve, we find that the total radiated energy is in excess of $10^{50}$\,erg, the progenitor star's pre-explosion mass-loss rate was $\ga 1 \times 10^{-2}\,{\rm  M_{\odot}~ yr}^{-1}$, and the total mass lost shortly before explosion was $\ga 1\,{\rm M_\odot}$, though the mass lost could have been considerably larger depending on the efficiency for the conversion of kinetic energy to radiation. The ultraviolet through near-infrared spectrum is characterised by two high density components, one with narrow high-ionisation lines, and one with broader low-ionisation H~I, He~I, [O~I], Mg~II, and Fe~II lines. The rich Fe~II spectrum is strongly affected by Ly$\alpha$ fluorescence, consistent with spectral modeling.  Both the Balmer and He~I lines indicate a decreasing CSM density during the late interaction period.  We find similarities to SN 1988Z, which shows a comparable change in spectrum at around the same time during its very slow decline.  These results suggest that, at long last, the shock interaction in SN~2005ip may finally be on the decline.
\end{abstract}


\section{Introduction}
\label{sec:intro}

Type IIn supernovae (SNe~IIn; see \citealt{filippenko97} and \citealt{smithHandbook} for reviews) are characterised by relatively narrow emission lines \citep{schlegel90} which are not associated with the SN explosion itself, but rather with dense circumstellar material (CSM) produced by pre-SN mass loss \citep{smith14all}.  Shock interaction and dust formation in the dense CSM often result in significant emission ranging from X-ray to radio wavelengths for many years post-explosion \citep[e.g.,][]{chevalier17,fox11, fox13}.  

SN 2005ip, discovered in NGC 2906 \citep{boles05} on 2005 November 5.163 (UT dates are used throughout this paper), is one of the more well-studied Type IIn explosions given its proximity to Earth ($\sim35$\,Mpc) and the fact that it has remained detectable for nearly 15 years post-explosion.  \citet{fox09} first reported a 3\,yr near-infrared (NIR) light-curve plateau corresponding to newly formed dust in the cold, post-shock shell.  \citet{smith09ip} published an optical light curve that showed an initial linear (in mag\,d$^{-1}$) decline from peak, followed by a late-time plateau attributed to ongoing shock interaction with the dense CSM.  The late optical plateau matched the NIR plateau, and \citet{smith09ip} presented spectra that revealed signatures of strong ongoing CSM interaction, as well as signatures of dust formation in both the SN ejecta and the post-shock cold dense shell.  SN~2005ip was unusual in displaying very pronounced narrow coronal emission lines in its spectrum, indicating that clumpy CSM was being strongly irradiated by X-rays from the shock interaction \citep{smith09ip}. Coronal lines have also been seen in the Type IIn SNe 1995N \citep{fransson02}, 2006jd \citep{stritzinger12}, and SN 2010jl \citep{fransson14}.  \citet{fox10} obtained a {\it Spitzer Space Telescope} Infrared Spectrograph (IRS; \citealt{houck04}) spectrum of SN 2005ip (the only mid-IR (MIR) spectrum of a SN~IIn to date), which revealed the presence of a second, cooler dust component associated with a pre-existing dust shell radiatively heated by this ongoing CSM interaction.  \citet{bevan18} modeled the spectral line evolution and confirmed that a significant amount of dust can be explained by dust formation in the ejecta, and \citet{nielsen18} found the dust properties to be unlike those of Milky Way dust.

\citet{stritzinger12} continued to monitor SN 2005ip throughout $\sim5$\,yr post-explosion and showed that the optical and NIR light curves underwent little decline over that time.  \citet{katsuda14} reported that X-ray observations at $\sim6$\,yr post-explosion exhibit a significant decrease in flux compared to previous epochs, suggesting the forward shock had finally overtaken the dense CSM in which the SN exploded.  Most recently, \citet{smith17} observed a temporary resurgence in the H$\alpha$ luminosity, which they interpreted to be the result of the forward shock crashing into an additional dense shell located $\lesssim 0.05$\,pc away, consistent with the distant pre-existing dust shell indicated by MIR observations \citep{fox11}. Overall, \citet{smith17} showed that the spectral evolution and X-ray emission from the decade-long CSM interaction phase of SN~2005ip was almost identical to the late interaction seen in SN~1988Z, the prototypical SN~IIn with long-lasting CSM interaction. 

The nature of the progenitors of both SN~2005ip and the broader Type IIn subclass remains ambiguous.  The mass-loss rates of SNe~IIn derived using various techniques are in the range $10^{-4}$--$10^{-1}$\,\ml\ and the total CSM masses are several \msolar\ \citep[e.g.,][]{smith09ip,smith07gy,smith08tf,fox09,moriya13,ofek14,fransson14,katsuda14,smithHandbook}.  Galactic analogs with such mass-loss rates and H-rich winds include luminous red supergiants, yellow hypergiants, and luminous blue variables (LBVs), each of which present further questions of their own (see \citealt{smith14all} for a review).  To complicate the interpretation even more, \citet{habergham12} find that SNe~IIn, including SN 2005ip, do not trace the most active star formation in galaxies, suggesting they are not exclusively associated with the most massive stars.  \citet{smith15} note that extremely luminous LBVs themselves do not trace regions of recent massive star formation, and go on to explain this effect with a binary progenitor scenario. \citet{nomoto95} also stressed the importance of binary evolution in various types of SN progenitors including SNe IIn.  On the other hand, SNe~IIn share many properties with SN impostors \citep{smith11impostor}, and pre-existing dust shells are reminiscent of the impostors' pre-SN eruptions, such as SN 2009ip \citep[e.g.,][]{mauerhan13}, where the eruption was linked to an LBV progenitor \citep[e.g.,][]{smith10iplbv}.  \citet{taddia15} find that long-lasting SNe~IIn have similar host-galaxy metallicities as SN imposters, which may be produced by LBV outbursts and have traditionally been thought to arise from massive stars.

The pre-SN mass-loss history of SNe~IIn may hold some clues since it probes the latest stages of massive-star evolution.  Differences in wind speeds, densities, compositions, and asymmetries result in distinguishable observational behaviours.  Given the dense CSM associated with most SNe~IIn, X-rays from shock interaction are often absorbed, reprocessed, and re-emitted at UV (predominantly) and optical wavelengths, making these wavelengths optimal for tracing CSM interaction and, thereby, the progenitor's mass-loss history \citep[e.g.,][]{chevalier94,chevalier11,chevalier17}.  Furthermore, when combined with other wavelengths, the UV may offer a quantitative estimate of the nucleosynthesis in the core of the star, which helps to constrain the initial mass of the progenitor prior to mass loss \citep{fransson02,fransson05,fransson14}.

Here we present multiwavelength observations of SN 2005ip at very late epochs, including UV, optical, NIR, MIR, and radio.  These data track the SN light curve as it declines in all bands.  Following \citet{stritzinger12}, throughout this paper we assume that the distance to NGC 2906 (the host galaxy) is 34.9\,Mpc, and we adopt $E(B - V ) = 0.047$\,mag  as the reddening \citep{smith09ip}.  When noted, we correct our spectral energy distributions (SEDs) and spectra for this colour excess assuming $R_V = A_V / E(B - V ) = 3.1$ using the reddening law of \citet{cardelli89}. Section \ref{sec:obs} presents the observations, while \S \ref{sec:analysis} analyzes the light curve and spectral evolution.  In \S \ref{sec:composition} we take a closer look at the UV spectra.  Finally, \S \ref{sec:summary} provides a discussion and conclusion of the work.

\section{Observations}
\label{sec:obs}

\subsection{{\sl HST\/} Imaging}
\label{sec:hst}

Table \ref{tab:hst} summarises {\it Hubble Space Telescope (HST)} imaging of SN 2005ip.  We obtained individual images from the Mikulski Archive for Space Telescopes (MAST), so they have been processed through the standard pipeline at the Space Telescope Science Institute (STScI).  We obtained photometry from individual {\em flc} frames using Dolphot \citep{Dolphin16} with the following parameters: FitSky=3, RAper=8, and InterpPSFlib=1, and the TinyTim model point-spread functions (PSFs).

\begin{table*}
\centering
\caption{{\it HST} Imaging \label{tab:hst}}
\begin{tabular}{ l c c c c c c c c}
\hline
UT Date & Epoch & Program & PI & Instrument & Grating/Filter & Central Wavelength & Exposure & Magnitude\\
YYYMMDD & (d)  & (GO)                  &     &                     &                          & (\AA)                            & (s)              & \\
\hline
20081118 & 1109 & 10877 & Weidong Li & WFPC2 & F450W & 4556.00 & 800 & 19.58 (0.01)\\
20081118 & 1109 & 10877 & Weidong Li & WFPC2 & F675W & 6717.00	 & 360 & 17.30 (0.01)\\
20081118 & 1109 & 10877 & Weidong Li & WFPC2 & F555W & 5439.00 & 460 & 19.34 (0.01)\\
20081118 & 1109 & 10877 & Weidong Li & WFPC2 & F814W & 8012.00  & 700 & 19.05 (0.01)\\
20161029 & 4011 & 14668 & Filippenko & WFC3 & F336W & 3354.85 & 780 & 20.82 (0.02) \\
20161029 & 4011 & 14668 & Filippenko & WFC3 & F814W & 8048.10 & 710 & 21.31 (0.01)\\
20180111 & 4450 & 15166 & Filippenko & WFC3 & F336W & 3354.85 & 780 & 21.32 (0.02)\\
20180111 & 4450 & 15166 & Filippenko & WFC3 & F814W & 8048.10 & 710 & 21.60 (0.01)\\
\hline
\end{tabular}
\end{table*}

\subsection{HST/STIS}
\label{sec:hst_spectra}

SN 2005ip was observed twice with the {\it HST}/STIS as part of programs GO-13287 and GO-14598 (PI O. Fox), as summarised in Table \ref{tab:stis}.  The one-dimensional (1D) spectrum for each observation is extracted using the CALSTIS custom extraction software stistools.x1d.  The default extraction parameters for STIS are defined for an isolated point source. For both G140L and G230L the default extraction box width is 7 pixels and the background extraction box width is 5 pixels. 

\begin{table}
\centering
\caption{{\it HST}/STIS/MAMA Spectroscopy \label{tab:stis}}
\begin{tabular}{ l c c c c}
\hline
UT Date & Epoch & Program & Grating & Exposure \\
YYYYMMDD & (d) &  (GO)     &                & (s)              \\
\hline
\multirow{2}{*}{20140328} & \multirow{2}{*}{3065} & \multirow{2}{*}{13287} &  G140L & 4752 \\
 & & &  G230L & 2744 \\
\multirow{2}{*}{20171021} & \multirow{2}{*}{4368} & \multirow{2}{*}{14598} &  G140L & 17,100 \\
 & & &  G230L & 19,008 \\
\hline
\end{tabular}
\end{table}

\subsection{Warm {\it Spitzer}/IRAC Photometry}
\label{sec:spitz}

The Warm {\it Spitzer} Infrared Array Camera (IRAC) \citep{fazio04} obtained several epochs of data for SN 2005ip, summarised in Table \ref{tab2} and plotted in Figure \ref{fig_photometry}.  We downloaded coadded and calibrated Post Basic Calibrated Data ({\tt pbcd}) from the {\it Spitzer} Heritage Archive.  Standard aperture photometry was performed with a radius defined by a fixed multiple of the PSF full width at half-maximum intensity (FWHM).  Template subtraction is typically implemented to remove contributions from the underlying galaxy, but in this case, no {\it Spitzer} template exists.  Furthermore, due to the rapid flux variations of the underlying galaxy, a standard annulus does not allow for selection of pixels corresponding to a local background associated with the SN (e.g., \citealt{fox11,szalai19}).  Instead, we selected our own background region.  The standard deviation of the background variations is smaller than the measured noise in the aperture photometry, even at the latest epochs, and does not contribute substantially to our error bars.  Most of these data were recently published by \citet{szalai19}, but day 4667 is newly published here.  The photometry for epochs 948 and 2057 doesn't precisely match the photometry of \citet{fox10,fox11,fox13} because the analysis in those papers implemented slightly different routines consisting of either different sized apertures or PSF fitting photometry, but they are within the error bars.  All {\it Spitzer} photometry in this paper was calculated using aperture photometry with consistent aperture sizes.

For plotting purposes in Figure \ref{fig_photometry}, we derive the integrated {\it Spitzer} luminosity by fitting a simple dust-mass model to the two MIR data fluxes, similar to those described by \citet{fox11}.  In this case, we assume 0.1\,$\micron$ graphite grains given the lack of the $\sim 10$\,\micron\  silicate feature in the MIR spectrum \citep{fox10,williams15}.

\begin{table}
\centering
\caption{$Spitzer$ Photometry\label{tab2}}
\begin{tabular}{c c c c c}
\hline
JD $-$                     & Epoch & PID & 3.6\,\micron & 4.5\,\micron \\
2,450,000 & (d)  & &   \multicolumn{2}{c}{(10$^{17}$ erg s$^{-1}$ cm$^{-2}$ \AA$^{-1}$)}\\
\hline
4628 & 948 & 50256 & 13.62(0.31) & 10.96(0.21)\\
5737 & 2057 & 80023 & 6.44(0.20) & 5.87(0.15)\\
6476 & 2796 & 90174 & 3.36(0.15) & 3.02(0.11)\\
6845 & 3165 & 10139 & 2.68(0.14) & 2.32(0.09)\\
7229 & 3549 & 11053 & 2.22(0.13) & 1.80(0.08)\\
8347 & 4667 & 14098 & 1.56(0.11) & 1.05(0.07)\\
\hline
\end{tabular}
\end{table}

\begin{table*}
\centering
\caption{Ground-Based Optical Photometry \label{tab_phot}}
\begin{tabular}{ l c c c c c c c c c}
\hline
JD $-$ & Epoch &  $g'$ & $B$ & $V$ & $R$ & $r$ & $I$ &  $i$ & Instrument\\
2,450,000 &  (d) & \multicolumn{6}{c}{mag} \\ 
\hline
6310 & 2630  & --- & ---& --- & --- & 17.99 (0.1)  & --- & 17.95  (0.04)  & RATIR\\
6713 & 3033  & --- & ---& --- & --- & 18.82 (0.1)  & --- & 18.76  (0.01)  & CSP\\
6753 & 3073 & --- & ---& --- & --- & 18.42  (0.1)  & --- & 18.31  (0.06)  & RATIR\\ 
7008 & 3328  & --- & ---& --- & --- & 18.73  (0.1) &  --- & 18.76  (0.01) & CSP \\ 
8134 & 4455  & 21.54 (0.1) & ---& --- & --- & 20.80 (0.1) & --- & --- & LBT \\ 
8455 & 4776  & --- & $>$20.83 & $>$20.88 & 19.92--24.54 & --- & $>$19.68 & --- & Keck/LRIS \\ 
\hline
\end{tabular}
\end{table*}

\begin{table*}
\centering
\caption{Ground-Based NIR Photometry \label{tab_nirphot}}
\begin{tabular}{ l c c c c c c c}
\hline
JD $-$ & Epoch & $Z$ &  $Y$ &  $J$ & $H$ & $K$ & Instrument\\
2,450,000 &  (d) & \multicolumn{6}{c}{mag} \\ 
\hline
6310 & 2630  &  17.47  (0.01)  & --- &  16.27  (0.01)  & 15.65  (0.02)  & --- & RATIR\\
6753 & 3073  &  17.91  (0.01)  & --- & 16.77  (0.02) & 16.09  (0.02) & ---  & RATIR\\ 
7435 & 3755  & ---  & --- & 19.81 (0.1) & 18.91 (0.1) & 17.3 (0.1) & UKIRT \\ 
7451 & 3771 &  ---  & --- & --- & --- & 17.2 (0.1) & UKIRT \\ 
7483 & 3803  & --- & --- & --- & 18.96 (0.1) & 17.2 (0.1) & UKIRT \\ 
\hline
\end{tabular}
\end{table*}

\subsection{Optical and NIR Photometry}
\label{sec:phot}

Tables \ref{tab_phot} and \ref{tab_nirphot} list and Figure \ref{fig_photometry} plots the new optical and NIR photometry of SN 2005ip.  We include some data obtained with the Reionization And Transients InfraRed camera \citep[RATIR;][]{butler12, fox12} mounted on the 1.5\,m Johnson telescope at the Mexican Observatorio Astrono\'mico Nacional on Sierra San Pedro M\'artir in Baja California, M\'exico \citep{watson12}.  The data were reduced, coadded, and analysed using standard CCD and IR processing and aperture photometry techniques, utilising online astrometry programs {\tt SExtractor} and {\tt SWarp}\footnote{SExtractor and SWarp can be accessed from http://www.astromatic.net/software.}. 

We also present two epochs of $r$ and $i$ photometry obtained during 2014 at Las Campanas Observatory with the 2.5\,m du Pont telescope by the Carnegie Supernova Project \citep[CSP;][]{hamuy06}. These images were reduced following standard procedures. PSF photometry of the SN was computed in the natural system using the local sequence stars presented by \citet{stritzinger12}. The reported photometric uncertainties account for both instrumental and nightly zero-point errors.

Additional NIR photometry was obtained from the 3.8\,m United Kingdom Infrared Telescope (UKIRT) on Maunakea using WFCAM2. $JHK$ observations were pipeline reduced by the Cambridge Astronomical Survey Unit (CASU).  Aperture photometry was performed using the DAOPHOT package in {\tt IRAF}\footnote{IRAF: the Image Reduction and Analysis Facility is distributed by the National Optical Astronomy Observatory, which is operated by the Association of Universities for Research in Astronomy (AURA), Inc., under cooperative agreement with the US National Science Foundation (NSF).}. Uncertainties were calculated by adding in quadrature photon statistics and zero-point deviation of the standard stars for each epoch.

One epoch of $g\prime$ and $r\prime$ photometry was obtained with the Multi-Object Double Spectrograph (MODS; Byard $\&$ O’Brien 2000) on the Large Binocular Telescope (LBT) on 2018 January 16. The $3 \times 240$\,s images in each filter were reduced and stacked using standard {\tt IRAF} procedures, and zero points were calculated using Sloan Digital Sky Survey (SDSS) standard stars in the field.  Uncertainties were calculated in the same manner as for the UKIRT data.

All magnitudes were initially calculated in their respective telescopes natural system.  Although not every telescope has a published report detailing their system, they all follow similar techniques as CSP \citep{contreras10}.  All photometric calibration was then performed using field stars with reported fluxes in both 2MASS \citep{skrutskie06} and the SDSS Data Release 9 Catalogue \citep{ahn12}. Uncertainties are dominated by errors associated with catalog stars.  

The most complicated point in Figure \ref{fig_photometry} is the 2018 $R$-band photometry from Keck (day 4776), when the SN is quite faint and comparable in broad-band flux to the underlying H~II region.  Aperture photometry, which includes some of the underlying H~II region, yields a magnitude of 19.92, which we take to be the SN upper limit (Table \ref{tab_phot}).  We also obtained a final epoch of Keck $R$-band imaging in 2019.  Under the assumption that the SN had faded completely (although it likely hadn't), we can use the 2019 data as a template for subtraction from the 2018 data.  This yields a magnitude of 24.54, which we take to be the SN lower limit in 2018.  In reality, the actual SN flux on day 4776 is somewhere between these limits.


\subsection{Optical Spectroscopy}
\label{sec:spec}

Table \ref{tab_opt_spectra} lists and Figure \ref{fig_interaction} plots the new optical spectra of SN 2005ip.  We obtained some spectra with the Low Resolution Imaging Spectrometer \citep[LRIS;][]{oke95} mounted on the 10\,m Keck~I telescope and the DEep Imaging Multi-Object Spectrograph \citep[DEIMOS;][]{faber03} mounted on the 10\,m Keck~II telescope.  For the Keck/LRIS spectra, we observed with a 1\arcsec\ wide slit and used either the 600/4000 or 400/3400 grisms on the blue side and the 400/8500 grating on the red side.  This observing setup resulted in wavelength coverage from 3200--9200\,\AA\ and a typical resolution of 5--7\,\AA.  For the Keck/DEIMOS spectra, we observed with a 1\arcsec\ wide slit and the 1200/7500 grating.  This observing setup resulted in wavelength coverage from 4750--7400~\AA\ and a typical resolution of $\sim 3$\,\AA.  In both cases, we aligned the slit the parallactic angle to minimise differential light losses \citep{filippenko82}.  These spectra were reduced using standard techniques \citep[e.g.,][]{foley03,silverman12bsnip1}. Routine CCD processing and spectrum extraction were implemented using the optimal algorithm of \citet{horne86}.  We flux calibrated these spectra and removed telluric absorption lines using alagorithms defined by \citet{wade88} and \citet{matheson00}.

We also obtained three epochs of spectroscopy with the Bluechannel (BC) spectrograph on the 6.5\,m Multiple Mirror Telescope (MMT) using the 1200\,l\,mm$^{-1}$ grating centred at 6300\,\AA\ \citep[see][]{smith17}.  We performed standard reductions, including bias subtraction, flat-fielding, and optimal spectral extraction.  We flux calibrated these spectra using spectrophotometric standards observed at similar airmasses.    


\begin{table}
\centering
\small
\caption{Ground-Based Spectroscopy \label{tab_opt_spectra}}
\begin{tabular}{ l c c c c c}
\hline
JD $-$ & Epoch & Instrument & Res. & Exp.\\
2,450,000 & (d) &  & (\AA) & (s) \\
\hline
4584 & 905 & Keck/LRIS & $\sim$9 & 1200\\ 
6246 & 2567 & Keck/DEIMOS & $\sim$3 & 2400\\ 
6778 & 3099 & Keck/LRIS & $\sim$6 & 1200\\   
7372 & 3693 & Keck/DEIMOS & $\sim$3 & 2400\\ 
7449 & 3770 & Keck/DEIMOS & $\sim$3 & 2400\\ 
7893 & 4214 & MMT/BC & $\sim$1 & 1200\\ 
8052 & 4373 & MMT/BC & $\sim$1 & 1200\\ 
8109 & 4430 & MMT/BC & $\sim$1 & 1200\\ 
8784 & 5105 & Keck/LRIS & $\sim$6 & 1200\\ 
\hline
\end{tabular}
\end{table}

\subsection{Chandra X-Ray Photometery}
\label{sec:chandra}

\begin{figure}
\centering
\includegraphics[width=3.3in]{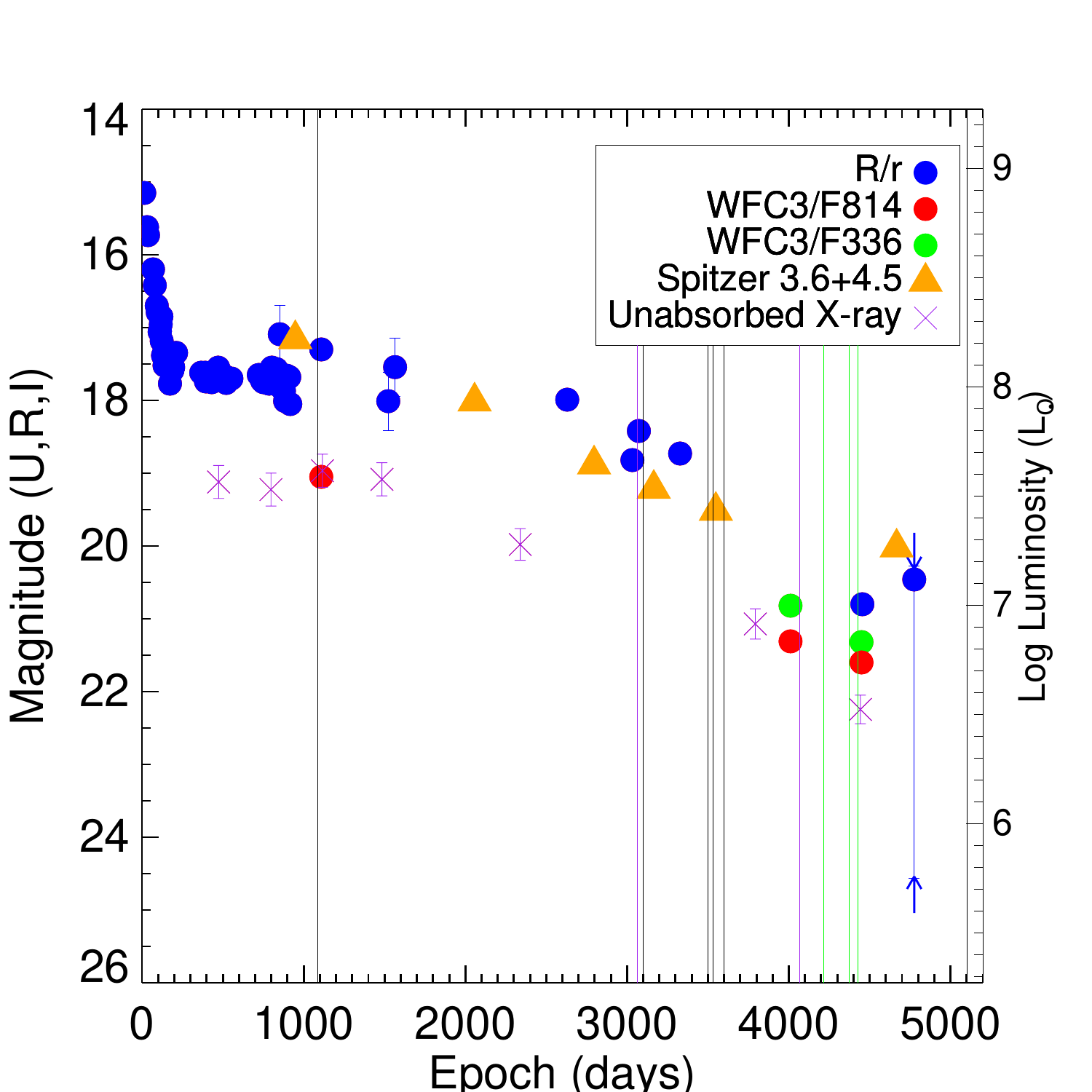}
\caption{Multiwavelength photometry of SN 2005ip, including data presented in this paper, \citet{smith09ip}, \citet{fox10}, \citet{stritzinger12}, \citet{katsuda14}, and \citet{szalai19}.  Vertical identify epochs on which the spectra presented here were obtained (black=Keck, green=MMT, and purple=Chandra).  Note that the left ordinate axis applies only to the $R/r$ band, F814, and F336 photometry, while the right-hand ordinate axis applies to the {\it Spitzer} integrated and X-ray luminosities.  The most complicated point in Figure \ref{fig_photometry} is the 2018 $R$-band photometry from Keck (day 4776), when the SN is quite faint and comparable in broad-band flux to the underlying H~II region.  Aperture photometry, which includes some of the underlying H~II region, yields a magnitude of 19.92, which we take to be the SN upper limit (Table \ref{tab_phot}).  We also obtained a final epoch of Keck $R$-band imaging in 2019.  Under the assumption that the SN had faded completely (although it likely hadn't), we can use the 2019 data as a template for subtraction from the 2018 data.  This yields a magnitude of 24.54, which we take to be the SN lower limit in 2018.  In reality, the actual SN flux on day 4776 is somewhere between these limits.}
\label{fig_photometry} 
\end{figure}

The \textit{Chandra X-ray Observatory} (CXO) Advanced CCD Imaging Spectrometer (ACIS; \citealt{garmire03}) observed SN 2005ip, summarised in Table \ref{tab:chandra} and plotted in Figure \ref{fig_photometry}.  As a reference, we also include details of the previous epoch of {\it CXO} observations from 2016 \citep{smith17}.  Similar to \citet{smith17}, we performed photometry and spectral extraction using the \texttt{specextract} package within the HEASOFT\footnote{http://heasarc.gsfc.nasa.gov/ftools} \textit{Ciao} software suite \citep{blackburn95}.  We model source and background spectra simultaneously using the \textit{Sherpa} package. For the source, we assume an absorbed single-temperature thermal plasma model (\textit{apec}) having solar abundances as defined by \citet{Asplund09}.  For the background, we assume a simple power law.  We set the equivalent neutral hydrogen column density in the interstellar medium (ISM) of $N_{\rm H} (\textrm{ISM})=3.7\times10^{20}$\,cm$^{-2}$ \citep{katsuda14}.  We also allowed for an additional intrinsic source of absorption for the SN.  The fits rely on a ${\chi}^2$ statistic with a Gehrels variance function.  In this case, we obtained a reduced ${\chi}^2$ value of 35 for 64 degrees of freedom.  We use these fits to derive photon energies in the range 0.5--8.0\,keV.  Compared to our previously reported \textit{Chandra}/ACIS observation on 2016 Apr. 3 \citep{smith17}, SN~2005ip exhibits a factor of $\sim 2$ reduction in the intrinsic flux. The temperature and self-absorption parameters of our thermal plasma model, however, have not changed significantly.

\begin{table*}
\centering
\caption{{\it Chandra}~Observations of SN 2005ip\label{tab:chandra}}
\begin{tabular}{c c c c c c c c c}
\hline
   Instrument &  JD $-$  & Days After & Exposure  & Counts  & $N_{\rm H}$ & $kT$ & 0.5--8\,keV & Luminosity \\
 & 2,450,000 & Outburst & (ks) & ($10^{-3}$  & ($10^{22}$ & (keV) &  Unabsorbed Flux ($10^{-13}$ & (10$^{40}$ erg s$^{-1}$)\\
& &  & & counts s$^{-1}$) & ${\rm cm^{-2}}$) & & erg s$^{-1}$ cm$^{-2}$) & \\
\hline
ACIS-S & 7482 & 3812 & 35.59 & 820 & $0.19_{-0.07}^{+0.08}$ & $5.0_{-0.9}^{+1.1}$ & $2.86_{-0.08}^{+0.12}$ & 3.2\\ 
ACIS-S & 8122 & 4453 & 41.22 & 619 & $0.11_{-0.11}^{+0.25}$ & $3.2_{-1.2}^{+2.0}$ & $1.2_{-1.1}^{+1.0}$ & 1.3\\ 
\hline
\end{tabular}
\end{table*}

\section{Analysis}
\label{sec:analysis}

\subsection{Light-Curve Evolution, Bolometric Luminosity, and Radiated Energy}
\label{sec:LC}
\begin{figure*}
\centering
\includegraphics[width=7.3in]{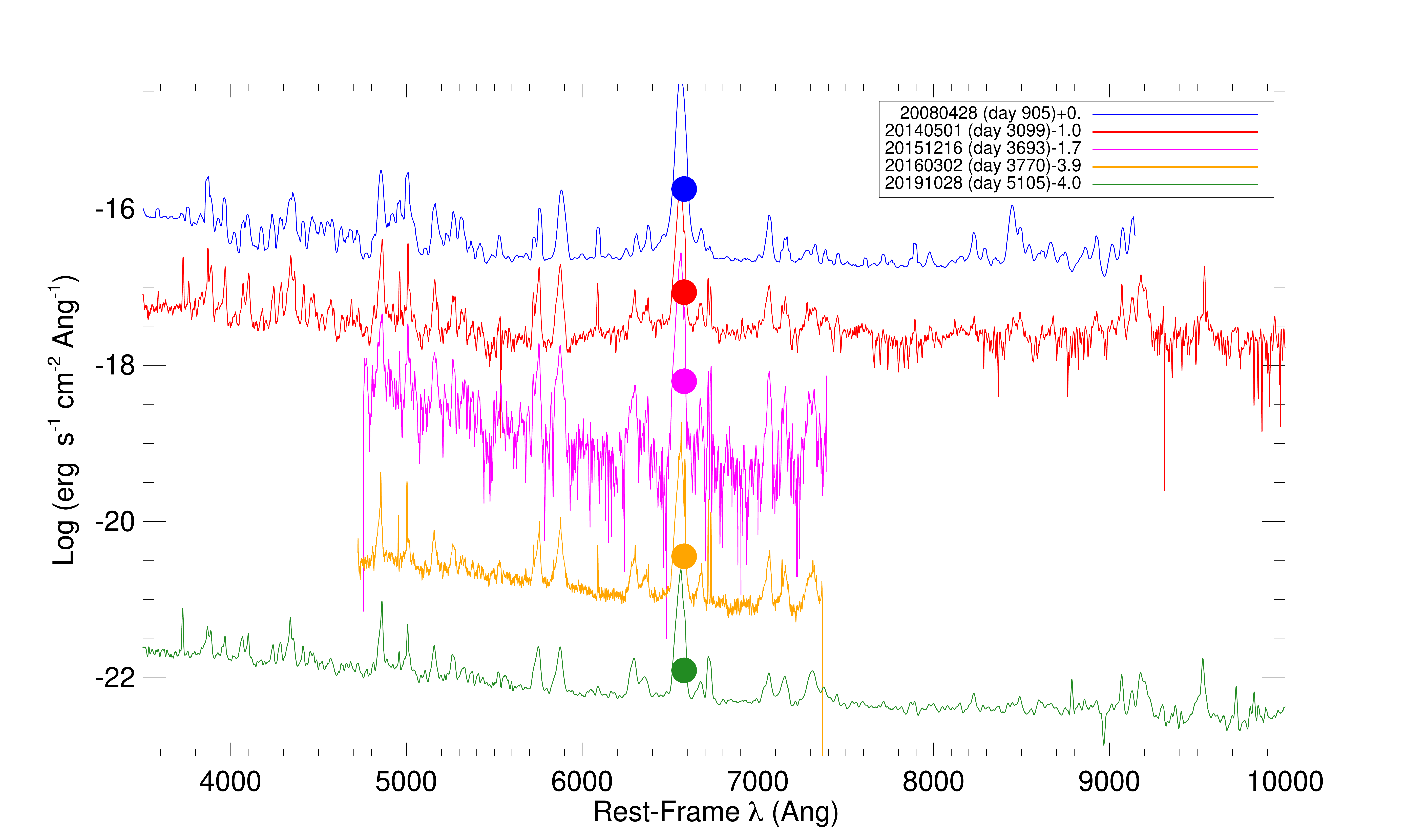}
\caption{Some of the ground-based optical spectra summarised in Table \ref{tab_opt_spectra}.  The spectra are scaled to the optical photometry for absolute comparisons at each epoch.}
\label{fig_interaction} 
\end{figure*}

Figure \ref{fig_photometry} shows decreasing fluxes at all wavelengths at days $\gtrsim 1500$ post-explosion, which is consistent with other SNe~IIn observed at such late epochs \citep{fox11}.  Based on the X-ray observations alone, \citet{katsuda14} attribute the decreasing flux to the forward shock having finally overtaken the dense CSM in which the SN exploded.  Figure \ref{fig_interaction} shows that most of the decreasing flux occurs in the strength of H$\alpha$ line, although the later spectra suggest that there may be some additional contribution from outside the H$\alpha$ line, possibly from a faint reflected light echo or a blend of very faint CSM interaction lines and their wings.

\begin{figure*}
\centering
\includegraphics[width=7in]{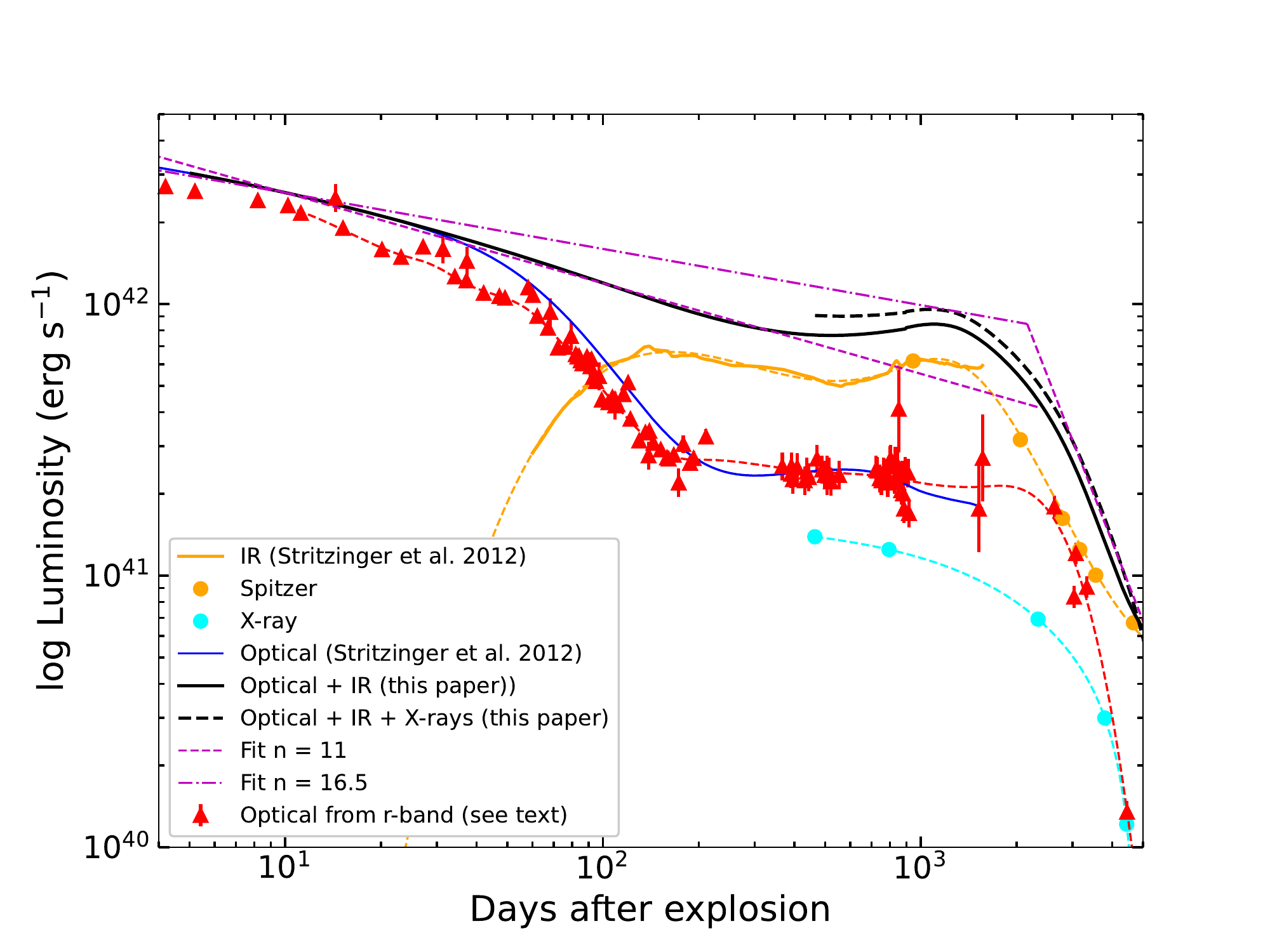}
\caption{Quasibolometric light curve of SN 2005ip. The solid lines are the optical (blue) and IR (orange) light curves from \citet{stritzinger12}, while the dashed lines are fits to the $r$-band (red) and {\it Spitzer} (orange) light curves, with photometry from this paper. The solid black line is the total constructed by merging the light curves from \citet{stritzinger12} and this paper. The dashed magenta lines show power-law fits to the two phases of the total luminosity (including and excluding X-rays, which are shown in cyan).}
\label{fig_bolom} 
\end{figure*}

We use the data from Figure \ref{fig_photometry} to construct a quasibolometric luminosity light curve (Figure \ref{fig_bolom}), which we will refer to simply as the bolometric light curve hereafter. For the optical, we use our $r$-band photometry and scale magnitudes to correspond to the optical luminosities from \citet{stritzinger12} in the range 350--900\,d, assuming that the spectrum does not change appreciably after this epoch. This includes both the ``hot" component and the lines in \citet{stritzinger12}, giving 
\begin{equation}
    L_{\rm optical} = 10^{-0.4 r + 48.45}.
\end{equation}

\noindent
The IR luminosities are described above. The MIR luminosities do not include the cold component discussed by \citet{fox10}, whose origin may be from more distant gas. Taken all together, the bolometric luminosity may be underestimated by at most 50\%. 
Because we do not include the far-IR, UV, or X-rays, this is likely to be a lower limit to the luminosity. The observed X-rays give an additional contribution of (15\%--20\%). Unfortunately, this component is not known before 460\,d.  For the total bolometric output only the observed X-ray luminosity should be included, not the fraction absorbed by the CSM, which is thermalised into UV and optical radiation. This fraction is most likely increasing for the earlier epochs, approaching 100\% as the column density of the CSM and ejecta ahead of the shock increases at early epochs. This is highly model dependent, and we do therefore not attempt to model it. To estimate the effect, however, we add the observed X-ray luminosity to the optical and IR contributions, shown as the dashed black line in Figure \ref{fig_bolom}.   

Figure \ref{fig_bolom} shows that the optical plus IR luminosity up to $\sim 700$\,d can be well described by a power law in time, $L \approx 2.6 \times 10^{42}\,(t/(10\,{\rm d}))^{-0.33}\,{\rm erg\,s^{-1}}$.  The power-law decay during the first phase can be well described by the similarity solution for a radiative shock in a CSM with a steady mass-loss rate \citep[e.g., ][]{chevalier17}: $L \propto v_{\rm s}^3 \propto t^{-3/(n-2)}$, where $n$ is the power-law index of the ejecta density profile.  If we use only the total optical plus IR light curve, the luminosity decrease corresponds to $n \approx 11$, somewhat steeper than found for SN 2010jl \citep{fransson14}.

 When we also include the X-rays, the dip in the optical plus IR light curve before the plateau partly fills in, increasing the luminosity by $\sim 20\%$. If we fit the luminosity from 10\,d to the peak of the ``bump" at $\sim 1000$\,d, we find  $L \approx 2.6 \times 10^{42}\,(t/(10\,{\rm d}))^{-0.21}\,{\rm erg\,s^{-1}}$, corresponding to $n=16.5$. Note, however, that this excludes any X-ray contribution at earlier epochs which would steepen the decline and decrease $n$.  In this context we  also note that a short-lived eruption, as is probably the case here, may have a density profile different from a steady $\rho \propto r^{-2}$ wind.  

After $\sim 2000$\,d the light curve breaks, and the decay is steeper, with $L \propto (t/(10\,{\rm d}))^{-3}$.  This behaviour signals the breakout of the shock wave from at least part of the dense CSM. Compared again to SN 2010jl, where the break occurred at $\sim 350$\,d, this is considerably later.

We integrate the bolometric light curve to estimate a total radiated energy of $1.7 \times 10^{50}$\,erg. We add $\sim 1 \times 10^{49}$\,erg from the X-rays. As we discuss above, however, this is most likely only a lower limit to the total radiated energy. 

Using the bolometric light curve we can also estimate the mass-loss rate.  Assuming a radiative shock in a steady wind, the total luminosity is given by 
\begin{equation}
L =  \frac{1}{2} \epsilon  \frac{\mdot}{v_{\rm w}}  v_{\rm shock}^3,
\label{eq_lum}
\end{equation}
where $\epsilon \leq 1$ is the efficiency for conversion,  $v_{\rm shock}$ is the shock velocity, $\mdot$ is the mass-loss rate, and  $v_{\rm w}$ is the wind velocity.  From the narrow high-ionisation lines we estimate the velocity of the pre-shocked CSM to be $\sim 100$\,km\,s$^{-1}$.  If we write the luminosity as $L(t)=L(t_*) (t/t_*)^\alpha$ and $v_{\rm shock}=v_{\rm shock}(t_0/t)^{1/(n-2)}$, we get
\begin{equation}
\mdot =  \frac{2 \  L(t_*) \ v_{\rm w}}{ \epsilon \ v_{\rm shock}(t_0)^3 }  \left( \frac{t_*}{t_0}\right)^{3/(n-2)}. 
\end{equation}

The shock velocity injects the largest uncertainty in the above equation. Because of electron scattering at early epochs, one cannot use the maximum wavelength shift of the blue wing. Figure 6 of \citet{smith09ip} shows that this line has a ``shoulder" at $\sim 5300$\,\kms\ on the blue side in the spectra at $\sim 100$\,d. As discussed in detail by \citet[][]{taddia20}, this shoulder may be caused by the macroscopic shock velocity, in contrast to the wings caused by electron scattering. Our late-time  H$\alpha$ spectra (Sec. \ref{sec:lprof})  indicate a velocity of $\sim 2700$\,\kms\ at  $\sim 1000$\,d, but the line profile still suggests contributions from electron scattering. Given this uncertainty, we will therefore scale the mass loss to a velocity of $3000\,{\rm km\,s^{-1}} $ at 1000\,d, as in Eq. \ref{eq:mdot}.

With $t_* = 10$\,d, $L(t_*)=2.6 \times 10^{42}\,{\rm erg\,s^{-1}}$, and $t_0=1000$\,d, we get
\begin{equation}
\begin{split}
\mdot = \frac{6.6 \times 10^{-3}}{ \epsilon } 
\left[\frac{v_{\rm w}} {100\,{\rm km\,s^{-1} }}\right]
\left[\frac{v_{\rm shock}(1000 \ {\rm d}) } {3000\,{\rm km\,s^{-1} }}\right]^{-3}  
\,M_{\odot}\,{\rm yr}^{-1}.
\end{split}
\label{eq:mdot}
\end{equation}
\noindent
The total mass of the SN 2005ip gas shell up to the break at $t_{\rm break}$ is then
\begin{equation}
\begin{split}
{\rm M_{tot}} = &  \frac{\mdot}{v_{\rm w}}  v_{\rm shock}(t_0) t_{\rm break} 
\left( \frac{t_0} { t_{\rm break} } \right)^{1/(n-2)}. 
\end{split}
\end{equation}
If we assume that the shock has exited the densest part of the CSM at $t_{\rm break} \approx 2000$\,d and with $\mdot$ from Eq. \ref{eq:mdot}, we get 
\begin{equation}
\begin{split}
{\rm M_{tot}} =   \frac{1.0}{\epsilon}  
\left[\frac{v_{\rm shock}(1000 \ {\rm d}) } {3000\,{\rm km\,s^{-1} }}\right]^{-2} 
 \left(\frac{t_{\rm break}}{2000\,{\rm d}}\right)^{8/9}\,M_\odot .
\end{split}
\label{eq:totm}
\end{equation}
\noindent
Equations \ref{eq:mdot} and \ref{eq:totm} together give a timescale of $\sim 150  \,(v_{\rm shock}(1000 \ {\rm d}) / 3000 \,{\rm km\,s^{-1}}) \, (v_{\rm w} /100\,{\rm km\,s}^{-1})^{-1}$~yr for the strong mass-loss episode.

If we instead use the flatter evolution of the luminosity in Figure \ref{fig_bolom} with $n=16.5$, the coefficient in Eq. \ref{eq:mdot} becomes $1.2 \times 10^{-2} \,{\rm M}_{\odot}\,{\rm yr}^{-1}$, and in Eq. \ref{eq:totm} the coefficient becomes $1.85  \,M_\odot$.

As seen from Eqs. \ref{eq:mdot} and  \ref{eq:totm}, besides the value of $v_{\rm shock}$, an important uncertainty in these estimates is the efficiency parameter, $\epsilon$, which depends on the importance of shock-wave instabilities and other multidimensional effects \cite[][]{taddia20}, and may be in the range $\sim 0.1$--1. For these reasons, the total radiated energy and total mass should be taken as lower limits. In addition, we only integrated the total mass swept to the break at $\sim 2000$\,d. Even if the light-curve steepening is caused by a decreasing density, the later evolution will certainly contribute to a substantial additional mass.

Other studies using X-ray and MIR data favour total CSM masses even higher than 10\,\msolar\ \citep{stritzinger12,katsuda14}.  These results are consistent with mass-loss rates that are all nearly $10^{-2}$\,\msolar\,yr$^{-1}$ for a period of several hundred years leading up to the progenitor's explosion, similar to what we find above.  The superluminous Type IIn SN 2010jl, for comparison, had a mass-loss rate of nearly $10^{-2}$\,\msolar\,yr$^{-1}$ and total mass loss of $\gtrsim3$~\msolar.

\subsection{Spectral Modeling and Line Identifications}

Basic modeling can be used to infer some qualitative estimates of the physical conditions in the CSM and SN, although  this should not be confused with a fully self-consistent spectral modeling \citep[e.g.,][]{dessart15}.  Already, there has been extensive discussion of the rich line spectra of SN 2005ip \citep{smith09ip, smith17, stritzinger12}.  The H~I, He~I, [N~II], [O~I], Mg~II, and [Ca~II] lines have FWHM $\approx 1500$\,\kms\ and are understood to originate from a denser, optically thick medium. We will refer to these as the low-ionisation component. In contrast,  the high-ionisation lines are all narrow, originating in the preshocked CSM with FWHM $\approx 350$\,\kms.  

\begin{figure*}
\centering
\includegraphics[width=7.5in]{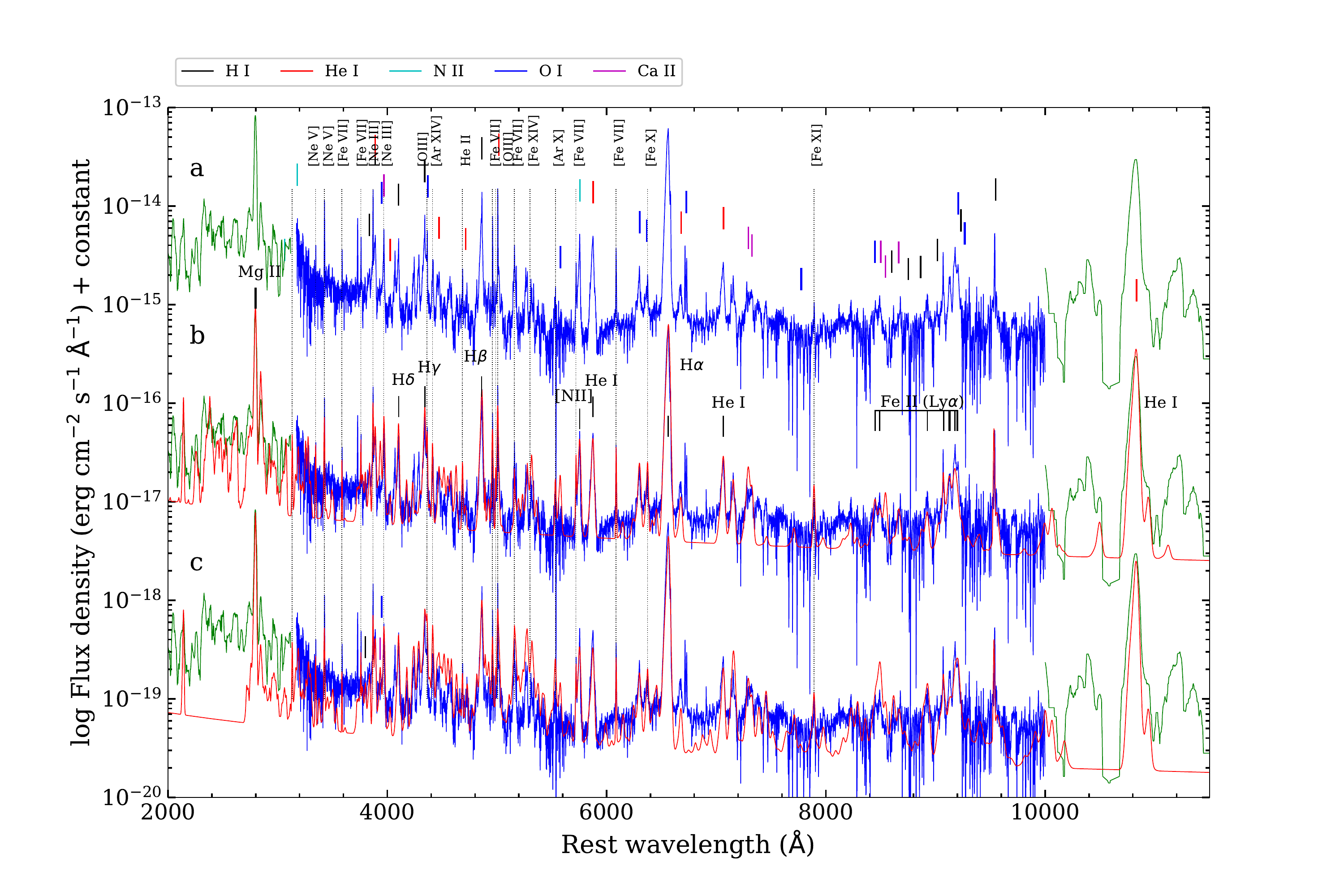}
\caption{Synthetic spectra with line identifications for the UV to MIR spectrum of SN 2005ip at 3065--3165\,d. The upper spectrum, (a), shows the observed data (green = near-UV and NIR, blue = optical) with colour-coded line identifications given in the legend above the figure, plus a number of marked high-ionisation lines, described in the text. The middle spectrum, (b), shows the same observed data together with the synthetic spectrum (red).  The Fe~II spectrum used is taken from the calculations by \citet{sigut03}. Spectrum (c) is the same, but with the the observed Fe~II spectrum of $\eta$ Carinae from \citet{zethson12}. Note especially the strong NIR Fe~II lines at $\sim 8450$--9200\,\AA, powered by fluorescence from Ly$\alpha$.
}
\label{fig:syntspec}
\end{figure*}

\subsubsection{Low-ionisation lines and Ly$\alpha$ fluorescence}
\label{sec_lowion}

As a tool for line identifications and for the diagnostics we have calculated a synthetic spectrum, including H~I, He~I, N~II, O~I, Mg~II, Ca~II, and Fe~II, which account for most of the spectral features. For H~I, He~I, N~II, O~I, and Fe~VII, we use multilevel model atoms, including collisional and radiative processes. We assume a two-zone model with temperature and density as parameters: one zone for the neutral and singly ionised elements and one zone for the high-ionisation ions.  We include optical-depth effects for the lines in the Sobolev approximation. The ionic abundances are treated as parameters.  The atomic data for high-ionisation stages are from the CHIANTI database \citep{landi12}. For H~I we use collision rates from \citet{anderson02} and for He~I from \citet{benjamin99}. Radiative transition rates and energy levels are mainly from NIST.

\begin{figure*}
\centering
\includegraphics[width=3.3in]{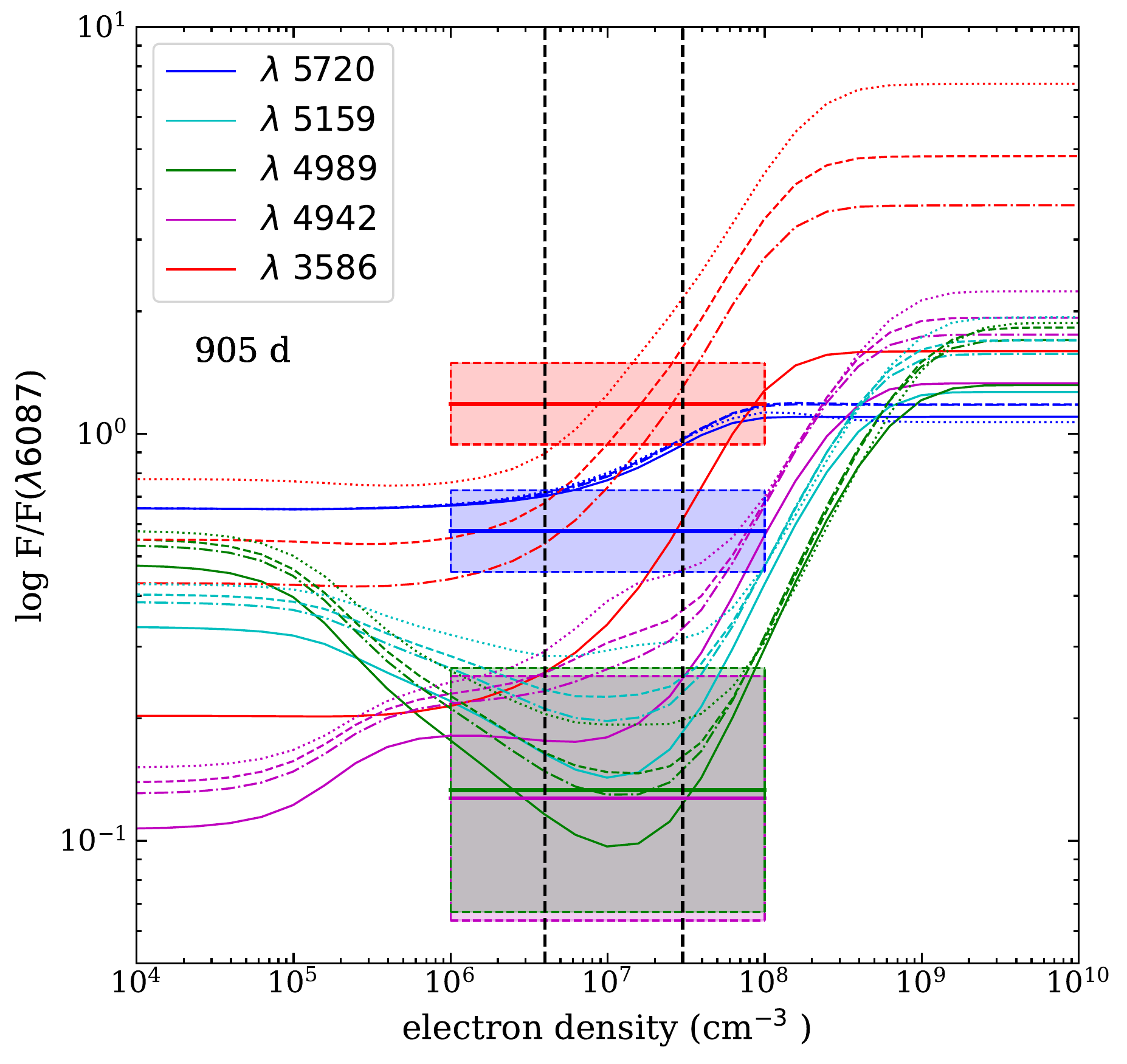}
\includegraphics[width=3.3in]{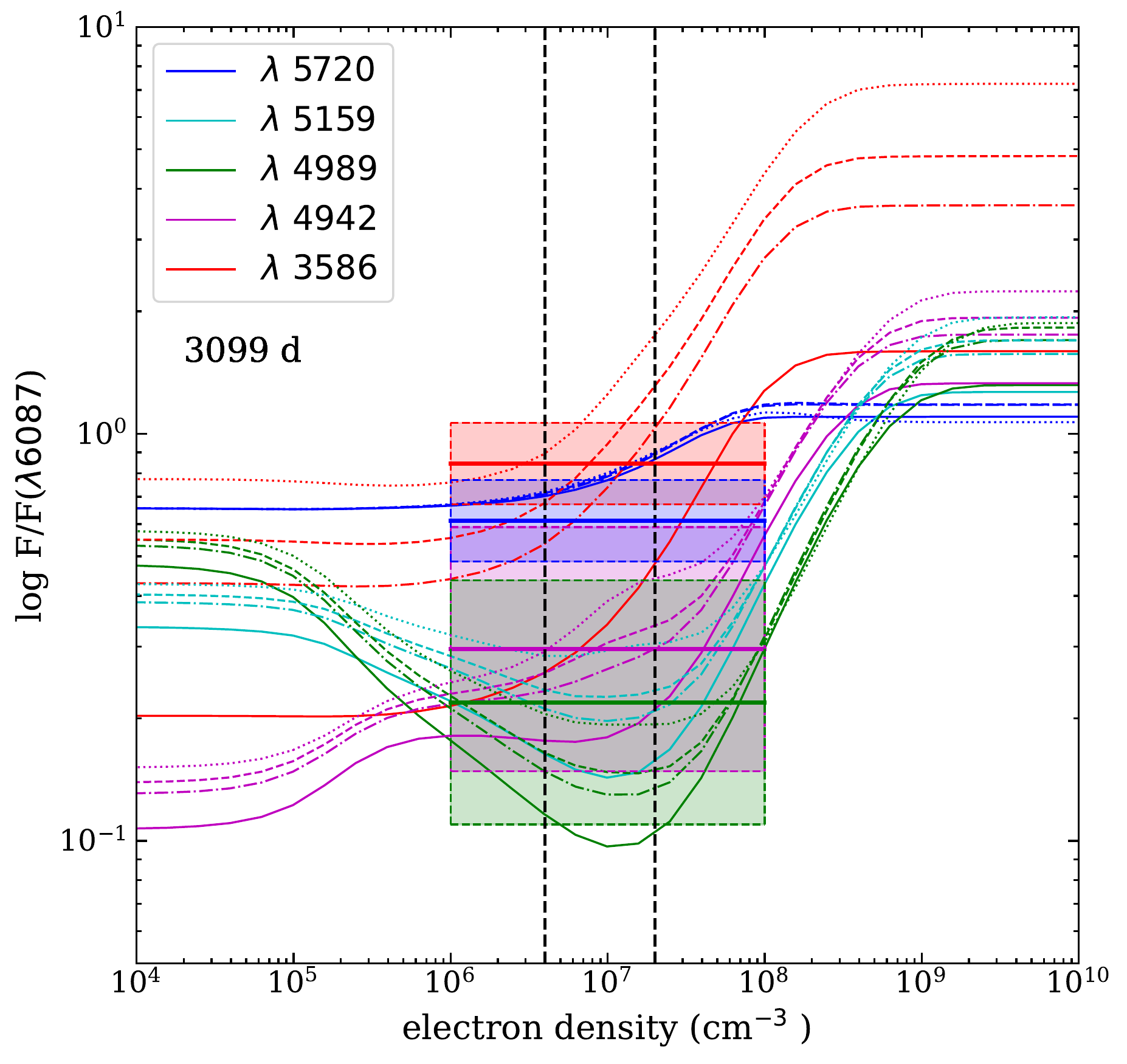}
\caption{Line ratios of the optical [Fe~VII] lines relative to the $\lambda 6086$ line as a function of density for 10,000 K (solid lines), 20,000 K (dot-dashed lines), 30,000 K (dashed), and 100,000 K (dotted). The left panel shows the observed ratios at 905\,d while the right is the same for 3099\,d. The vertical dashed lines shows the density range compatible with the different line ratios. The colour code of the modelled and observed ratios is given in the legend. The solid horizontal lines give the observed ratios, while the dashed lines indicate approximate error bars.}
\label{fig:fe7}
\end{figure*}

We have not attempted a similar calculation for Fe~II, in spite of the large number of lines in the spectrum. This requires a more detailed calculation, including radiative excitation by overlapping lines, leading to fluorescence through line coincidences.  While this is not a problem for the ions above, there are strong indications that the Fe~II spectrum is much affected by fluorescence. 
In particular, the prominent features at $\sim 8450$--9200\,\AA, not usually seen in SN spectra, are noteworthy. The peaks near 9200\,\AA\ do not coincide with any Paschen lines or Mg~II, both of which are detected at longer wavelengths. Instead, we argue that this emission arises from Fe~II lines powered by fluorescence.  

Excitation of Fe~II by fluorescence from Ly$\alpha$ was first discussed for cool and symbiotic stars  \citep{johansson84}, and later for active galactic nuclei (AGNs) and LBVs, in particular for $\eta$ Carinae \citep[see, e.g.,][for reviews]{johansson93,hartman13}. In the SN context fluorescence by Ly$\alpha$ was found to be important for the Type IIn SN 1995N \citep{fransson02}.  The most important branch is pumping by Ly$\alpha$, primarily from the $a^4D^e$ excited level at 1.04\,eV in Fe~II to levels $\sim 11.2$\,eV above the ground state \citep{sigut98,sigut03}. The cascade from these levels results in NIR lines at $\sim 8450$--9200\,\AA\ and UV lines at $\sim 2300$--2900\,\AA.  While both the UV lines and optically forbidden lines may also be excited by thermal collisions, the NIR lines require a very large excitation energy and are characteristic signatures of Ly$\alpha$ pumping. 

\begin{figure*}
\centering
\includegraphics[width=6.7in]{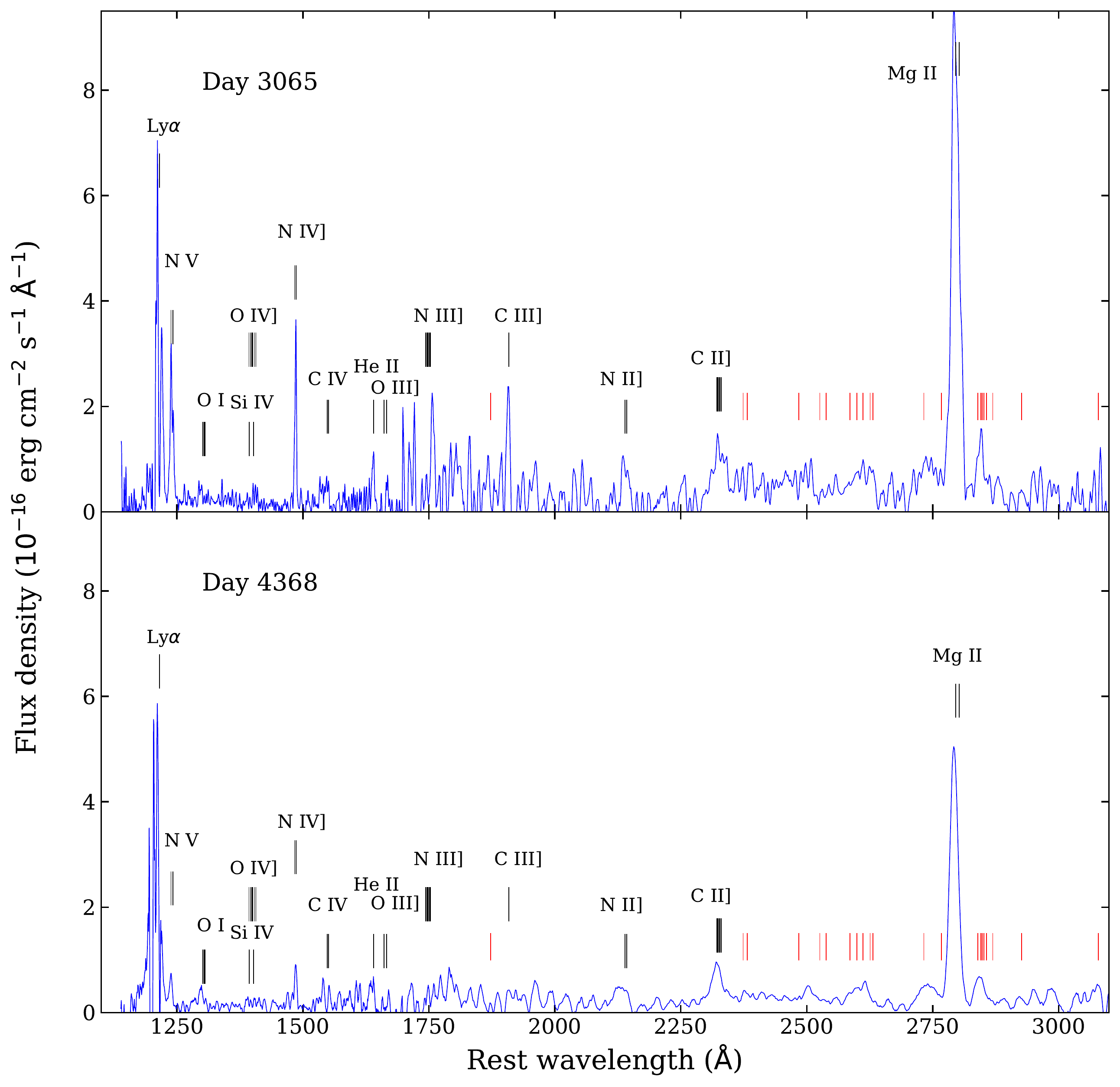} \\
\caption{{\it HST}/STIS spectra of SN 2005ip on days 3065 and 4368 post-explosion.  No extinction correction or background subtraction is applied. The wavelengths of probable line identifications are shown. The red vertical bars mark the wavelengths of the strongest Fe~II lines from the list of \citet{sigut03}. }
\label{fig_stis_spectrum} 
\end{figure*}

In SN 2005ip, the strong Ly$\alpha$ line reaches to $\sim 15$\,\AA\ on the red side, and more on the blue, although the blue is contaminated by the geocoronal Ly$\alpha$ and interstellar absorption (Figure \ref{fig_stis_spectrum}). \citet{sigut03} find 15 transitions from the $a^4D^e$ level within $\pm 3$\,\AA\ of Ly$\alpha$, so pumping can occur in a large number of transitions. Pumping from other low levels of Fe~II may occur.

To model the Fe~II spectrum we have therefore taken two approaches. In one, we have used the relative intensities by \citet{sigut03}, based on theoretical calculations. These are tuned for typical AGN conditions and may therefore give somewhat different intensities from those expected in the CSM of a SN. In particular, the velocity field is very different and nonlocal scattering may be important, resulting in other pumping channels. The qualitative results should, however, be similar. In the other approach we use the observed UV to NIR spectrum of $\eta$ Carinae by \citet{zethson12}.  The line intensities are convolved with Gaussian profiles and we add a continuum with $F_\lambda \propto \lambda$. 

Comparing the spectra in Figure \ref{fig_interaction} we see little evolution between day 905 and at least to day 3770. Because the spectrum around 3100 days has a coverage in both the UV and the NIR we will here concentrate on this spectrum.   We return to the other spectra and changes between these below.

Figure \ref{fig:syntspec} shows the result of a ``best-fit" calculation for the two Fe~II line cases mentioned above.  For both models, we note a steep Balmer decrement, with $F({\rm H}\alpha)/F({\rm H}\beta) \approx 9$. This ratio is comparable to that of other SNe~IIn \citep{fransson14} and is mainly a result of the large optical depth of H$\alpha$. This situation, often referred to as ``Case C," is discussed in detail by \cite{xu92}. For the higher-order Balmer lines, as well as the Paschen lines in the NIR, there is good agreement with the observations for this model. Similar agreement is found for the He~I lines, where the comparatively high $\lambda 7065/\lambda 5876$ ratio, as well as the  $\lambda 10930/\lambda 5876$ ratio, are a result of high optical depth in these lines \citep[see][]{karamehmetoglu19}.

The Fe~II emission shows a prominent line feature in the $\sim 8450$--9200\,\AA\ range, with the three peaks at 9076, 9126, and 9177\,\AA, which are well reproduced by both models. The lines at 8451\,\AA\ and 8490\,\AA\ also agree well with the simulations, confirming the contribution of Ly$\alpha$ fluorescence. The feature at $\sim 2845$\,\AA\ is most likely to be Fe~II rather than Mg~I $\lambda\lambda 2852$, 2857.  This line complex consists of a number of strong lines from an upper $d^4D$ level fed by transitions from the $v^4F$ level, which also gives rise to the $\sim 9175$\,\AA\ peak. The simulations also show the very large number of Fe~II lines in the region 4000--6000\,\AA, which makes an unambiguous identification of other weak lines challenging. 

The \citet{sigut03} model yields intensities that agree well at most wavelengths, although the $\sim 2848$\,\AA\  peak is overproduced by a factor of $\sim 3$.  The $\eta$ Carinae spectrum gives better agreement with the optical range 4000--5500\,\AA, while the AGN simulation agrees better with the fluorescence features in the NIR and UV.  The relative line intensities depend on both the atomic data, especially collisional, and the physical conditions.  The AGN environment, for example, has a density of $\sim 4\times 10^9\,{\rm cm}^{-3}$, which is much higher than that of the SN CSM. The calculated intensities also depend on the continuum level, which can be quite uncertain. Regardless of model, however, we find that there is strong evidence for the importance of Ly$\alpha$ fluorescence in the spectrum of SN 2005ip. 

\subsubsection{High-ionisation CSM lines}
\label{sec_highion}
  
High-ionisation ions were observed early \citep{smith09ip} and grew stronger with time \citep{smith17}.  Figure \ref{fig:syntspec} shows that in our most recent spectra, we identify a large number of these high-ionisation lines, including He~II $\lambda 4685.6$,  [O~III] $\lambda\lambda 4363.2$, 4958.9, 5006.8, [Ne~III] $\lambda\lambda 3869.1$, 3967.5, [Ne~V] $\lambda\lambda 3345.8$, 3425.5, [Fe~VII] $\lambda\lambda 3586.3$, 3758.9, 4942.5, 4988.6, 5158.4, 5720.7, 6087.0, [Fe~X] $\lambda 6374.5$, and [Fe~XI] $\lambda 7891.8$. In addition, [Ar~X] $\lambda 5533$, [Ar~XIV] $\lambda 4412.3$, and [Fe XIV] $\lambda5302.9$ may be present, although likely blended with [Fe~II] lines at later epochs.  

The numerous [Fe~VII] lines offer a diagnostic of the density and temperature from the region where these arise. The FWHM of these lines is $\sim 350$\,\kms, similar to the other high-ionisation lines.  The main uncertainty of the line fluxes comes from blending and the continuum level.  The continuum uncertainty affects the $\lambda\lambda 4988.6$, 5158.4 lines most.  The [Fe~VII] $\lambda 6087.0$ line is an important diagnostic, but nearly coincides with the $\lambda 6086.37$ line of [Ca~V]. From their transition rates, [Ca~V] $\lambda 6086.37$ should have an intensity $\sim 0.19$ of [Ca~V]  $\lambda 5309.1$. The latter is a blend, but using its peak intensity as an upper limit to the flux and a ratio of the $\lambda 6086 / \lambda 5309$ features of $\sim 2.4$, we predict a maximum contribution from [Ca~V] to the  $\lambda 6086$ line of 8\%. We therefore conclude that this line is mainly due to [Fe~VII]. The [Fe~VII] $\lambda\lambda 3586.3$, 3758.9 lines have a common upper level and their ratio is therefore fixed by their transition probabilities to $1:1.5$, which agrees with the observed ratio, and we therefore only discuss the $\lambda\lambda 3586.3$ line. 

From the observed spectrum at 905\,d, $F(\lambda 3586)/F(\lambda 6087) \approx 1.08$ and  $F(\lambda 5721)/F(\lambda 6087) \approx 0.57$. At 3399,d the corresponding ratios are  $F(\lambda 3586)/F(\lambda 6087) \approx 0.84$  and  $F(\lambda 5721)/F(\lambda 6087) \approx 0.61$. The uncertainties in these ratios  mainly come from the assumed reddening and the continuum level, and we estimate these to be $\sim 30$\%. Because of blending, the other ratios have considerably larger estimated uncertainties.

Figure \ref{fig:fe7} plots the theoretical line ratios for the strongest lines as a function of the electron density for temperatures from $10^4$\,K to $10^5$\,K.  Collision strengths are obtained from \cite{berrington00} and transition rates are obtained from NIST with a 9-level model atom. All ratios are relative to the strong $\lambda 6087.0$ line.  Overplotted are the measured line ratios from days 905 and 3099.

We place the most emphasis on the strongest lines, represented by the $F(\lambda 3586)/F(\lambda 6087)$ and  $F(\lambda 5721)/F(\lambda 6087)$ ratios. The strong temperature dependence of the $F(\lambda 3586)/F(\lambda 6087)$ ratio, together with the $F(\lambda 5721)/F(\lambda 6087)$ ratio, points to a temperature $\gtrsim 3\times 10^4$\,K on day 905 and $\gtrsim 2\times 10^4$\,K on day 3099. Day 905 may have a temperature as high as $\sim 10^5$\,K.  Photoionisation calculations for X-ray-illuminated plasmas by \citet[][]{kallman82} (e.g., their Model 3) show that Fe~VII is most abundant at $\sim 3 \times 10^4$\,K, consistent with these observations. For comparison, Fe~XIV, which is seen in at least the earlier spectra, arises at $\sim 6\times 10^4$\,K.

The electron density is $\sim 4\times 10^6$--$3 \times 10^7\,{\rm cm}^{-3} $ for day 905, and a similar density range, $\sim 4\times 10^6$--$2 \times 10^7\,{\rm cm}^{-3}$, for day 3099.  It is difficult to draw any strong conclusions about a changing density between these epochs, but a decrease should have been expected. As pointed out by \citet{berrington00}, there is likely considerable uncertainty in the atomic data, adding to the uncertainty in the observed fluxes.  However, even considering this, it is quite remarkable that the SN is interacting with a CSM with density $\gtrsim 4 \times 10^6\,{\rm cm}^{-3}$ at $\gtrsim 3000$\,d after explosion. 

The density we find here can be compared to the density inferred from the bolometric light curve in Sec. \ref{sec:LC}. With a mass-loss rate $\mdot \approx 1 \times 10^{-2}\,\epsilon^{-1}\,{\rm  M_{\odot}\,yr}^{-1}$ and  velocity $\sim 3000$\,\kms\  at $\sim 1000$\,d, one obtains a density $\sim 4.5 \times 10^6 \epsilon^{-1}\,{\rm cm}^{-3}$ at 1000\,d. While this is compatible with the range we find above from the forbidden lines, one should note that the density from the bolometric light curve depends on $\epsilon$ and is therefore only a lower limit; it may be up to an order of magnitude higher. We also note that the forbidden lines are not affected by electron scattering and should therefore arise in a region different from that responsible for the bulk of the radiation, close to the shock. A natural scenario is therefore that the forbidden, high-ionisation lines arise outside the region close to the shock which is optically thick to electron scattering, and where the Balmer lines and most of the radiation originate.

\subsubsection{The UV spectrum}
\label{sec:composition}

 Figure \ref{fig_stis_spectrum} shows the extracted UV spectra (no background subtraction is applied) and
 identifies the UV spectral lines, including the strong presence of Ly$\alpha$, \ion{N}{V} $\lambda\lambda$1238.8, 1242.8, \ion{N}{IV} $\lambda\lambda$1483.3, 1486.5, \ion{N}{III}] $\lambda\lambda$1746.8--1754.0, \ion{C}{III}] $\lambda\lambda$1906.7, 1908.7, and \ion{Mg}{II} $\lambda\lambda$2795.5, 2802.7. Weaker lines also exist, specifically \ion{He}{II} $\lambda$1640, N~II] $\lambda\lambda$2139.0, 2142.8, and \ion{C}{II}] $\lambda\lambda$2323.5--2328.1. We identify the feature at $\sim 2600$\,\AA\ with the Fe~II $4 s a^{6}D$ to $4 p z^{6}D^{0}$ multiplet. There may be additional blends at $\sim 2500$\,\AA\ and $\sim 2750$\,\AA, but their signal-to-noise ratio (S/N) is lower.  We note that the \ion{N}{III}] $\lambda\lambda$1746.8--1754.0 multiplet is redshifted to $\sim 1758$\,\AA.  

Table \ref{tab_uvlineflux} lists the reddening-corrected fluxes for the strongest UV lines from day 3065 shown in Figure \ref{fig_stis_spectrum} (with the background level subtracted). Several caveats should be noted.  The low S/N in the region above 1700\,\AA\ makes the continuum difficult to estimate, so the systematic error is approximated by comparing different wavelength ranges from the lines. The flux of \ion{C}{IV} $\lambda\lambda$1548, 1551 is likely underestimated owing to the presence of both the Galactic and host-galaxy ISM absorption.  The \ion{O}{III}] $\lambda$1664 line is weak and in a noisy region of the spectrum, so the calculated flux should only be considered an upper limit. 

\begin{table}
\centering
\caption{Reddening-corrected fluxes of strong UV lines from day 3065. \label{tab_uvlineflux}}
\begin{tabular}{ l l c c}
\hline
\hline
Species                    & \phantom{} \phantom{}Wavelength & Flux & Uncertainty\\
	&\phantom{} \phantom{} \phantom{} \phantom{} \phantom{} \phantom{} (\AA)  &  \multicolumn{2}{c}{(10$^{-15}$\,erg\,s$^{-1}$\,cm$^{-2}$)}\\
\hline
\ion{N}{V}		&	1238.8, 1242.8		&	4.80		&	0.07	\\
\ion{N}{IV}]	&	1483.3, 1486.5		&	2.72		&	0.15	\\
\ion{C}{IV}		&	1548.2, 1550.7		&	1.14		&	0.11	\\
\ion{He}{II}	&	1640.4			&	0.88		&	0.26	\\
\ion{O}{III}]	&	1660.8, 1666.2		&	$<$0.5	&	--	\\
\ion{N}{III}]	&	1746.8--1754.0		&	3.01		&	0.35\\
\ion{C}{III}]	& 	1906.7, 1908.7		&	2.55		&	0.34\\
\end{tabular}
\end{table}

An estimate of the CSM density for the first epoch can be obtained from the ratio of the \ion{N}{IV}] $\lambda$1483.3 and $\lambda$1486.5 lines, which we calculate to be $\lambda  1483.3/\lambda 1486.5 \approx 0.3$. For an X-ray-ionised plasma the N~IV abundance peaks where the temperature is (2--3) $\times 10^4$\,K \citep{kallman82}.  Most important, the line ratio is sensitive to density but relatively insensitive to temperature.  From Figure 1 of \citet{keenan95}, the observed $\lambda  1483.3/\lambda 1486.5$ ratio corresponds to an electron density of (2.0--3.6) $\times 10^5$\,cm$^{-3}$ in the temperature range (1--2)  $\times 10^4$\,K.  We conclude that the CSM density should be safely lower than the critical densities of the semiforbidden lines.

\subsubsection{Line profiles}
\label{sec:lprof}
\begin{figure}
\centering
\includegraphics[width=3.3in]{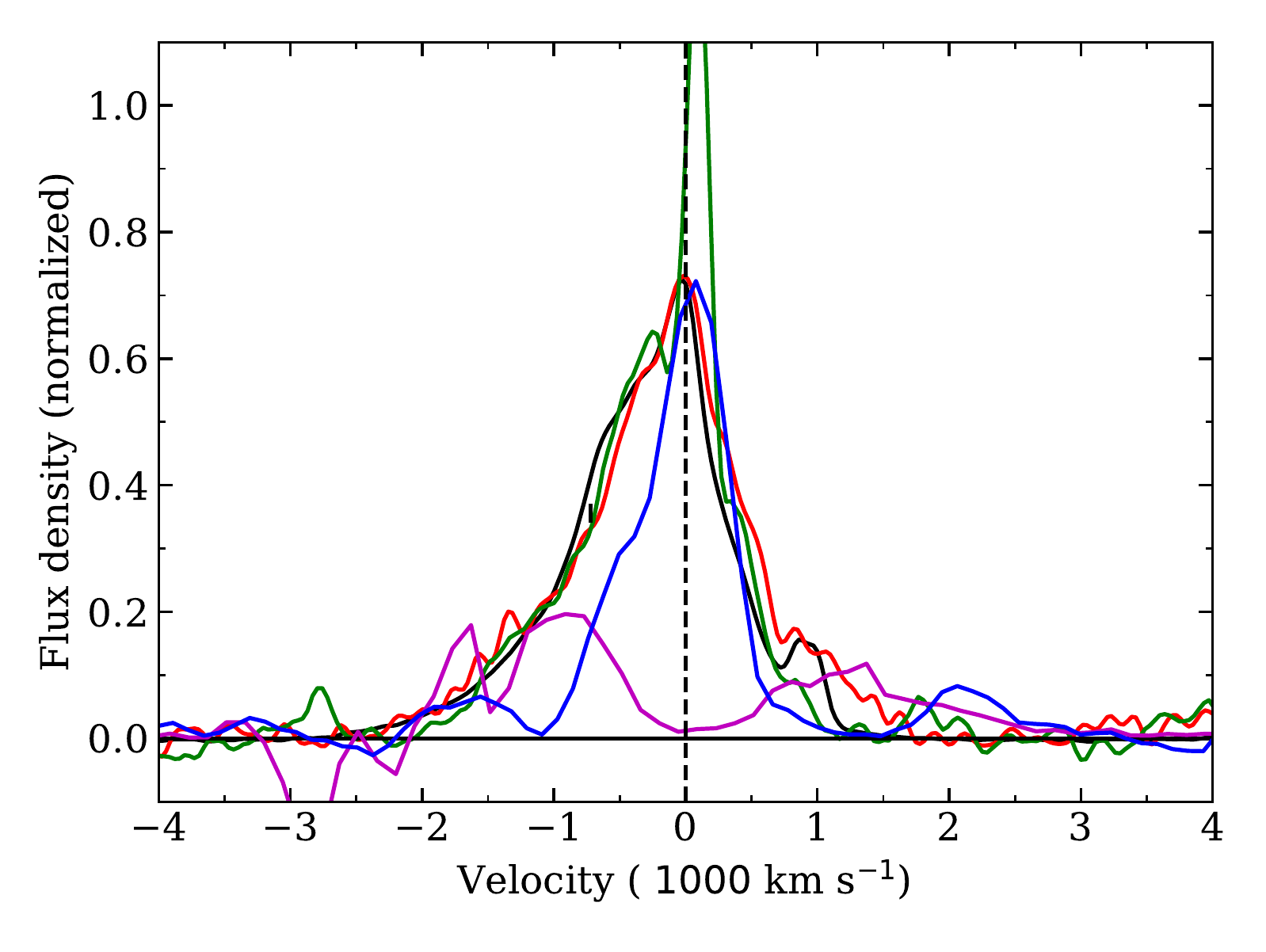}
\caption{Line profiles of H$\alpha$ (black), H$\beta$ (green), \ion{He}{I} $\lambda$5876 (red), Ly$\alpha$ (magenta), and \ion{N}{IV}] $\lambda\lambda$1483.3, 1486.5 (blue). Note the narrower line profile of the \ion{N}{IV}] lines, as well as the contribution of the semi-forbidden $\lambda$1483.3 component on the blue side.}
\label{fig_halpha} 
\end{figure}

Figure \ref{fig_halpha} compares the velocity profile of several UV and optical lines at $\sim 3100$ days.  The H$\alpha$ line extends to $\sim 2600$\,km\,s$^{-1}$ on the blue side and only $\sim 1000$\,km\,s$^{-1}$ on the red. H$\beta$, \ion{He}{I} $\lambda$5876, and \ion{Mg}{II} $\lambda\lambda$2796, 2803 have similar line profiles.  While the blue side can be well fit with an electron-scattering wing, the line centroid is shifted to the blue, with a ``shoulder" at $\sim -800\,\kms$. As shown by \citet{taddia20} and \citet{dessart15}, this type of profile can be explained by a combination of emission from the shock wave and emission from the pre-shock CSM. The emission from the shock is responsible for the blueshifted shoulder, while the central component is coming from pre-ionised gas in the CSM. Both components are affected by electron scattering in the low-velocity CSM. This gives rise to the smooth, blue wing shortward of the ``shoulder." Because of the smoothing by the electron scattering, the velocity of the ``shoulder" is lower than the shock velocity.

\begin{figure*}
\centering
\includegraphics[width=7in]{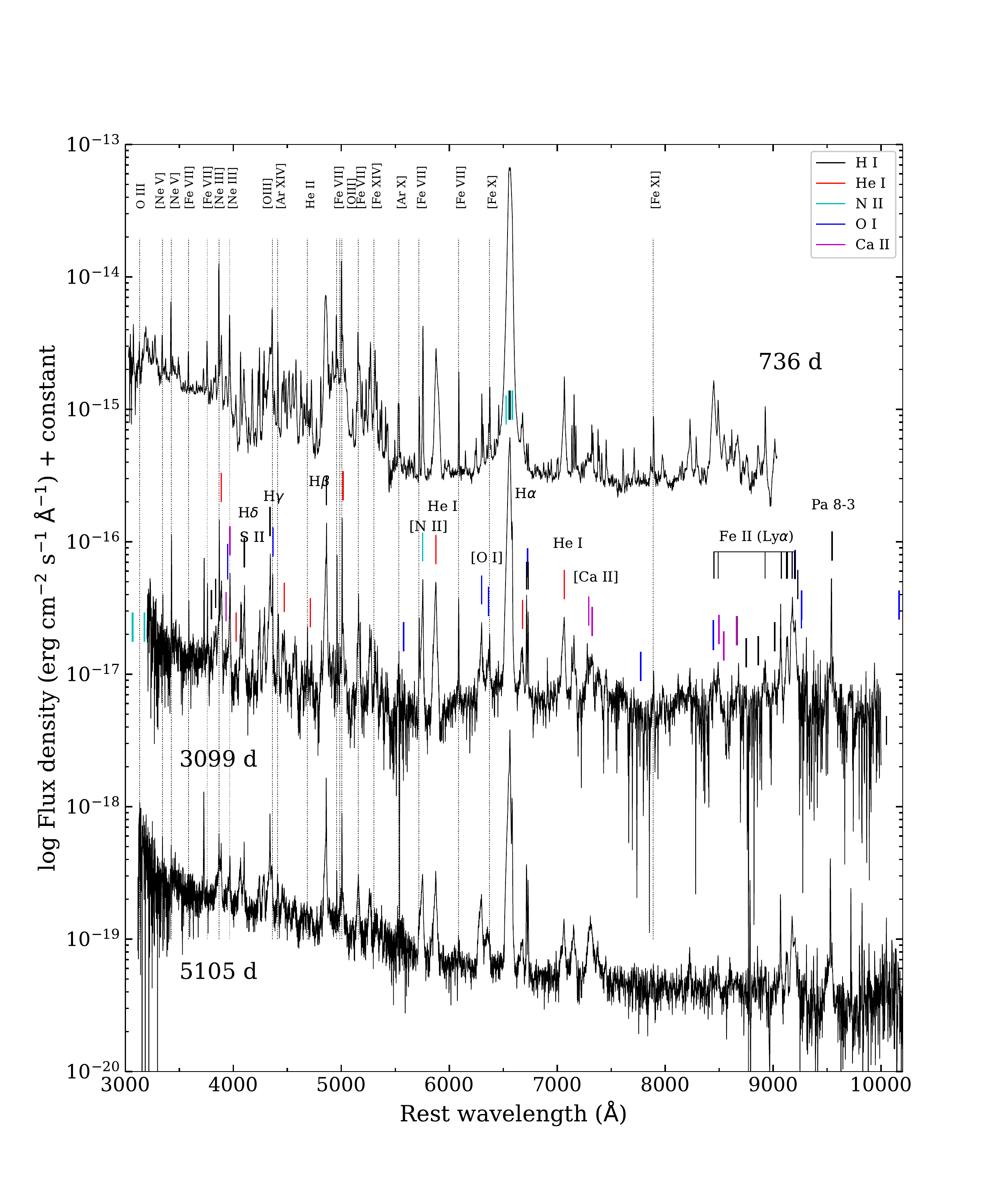}
\caption{Evolution of the optical spectrum from day 736 to day 5105 with identification of the high-ionisation lines at the top and colour coding and identification of the low-ionisation lines for the day 3099 spectrum.  }
\label{fig:spec_evol}
\end{figure*}

Both H$\beta$ and He~I $\lambda 5876$ have line profiles similar to that of H$\alpha$, as expected if these arise from similar regions (Fig. \ref{fig_halpha}). The Ly$\alpha$ line has a profile broadly consistent with H$\alpha$, although strongly distorted by a central absorption feature from both the Galactic and host-galaxy ISM.  

The \ion{C}{III}], \ion{N}{III}], \ion{N}{IV}], and \ion{N}{V} UV lines are considerably narrower than H$\alpha$, and consistent with being unresolved at the  $\sim 300$\,km\,s$^{-1}$ instrumental resolution of STIS. These lines do not show the velocity shift of the Balmer and He~I lines, and most likely come from the same optically thin, highly ionised region of the CSM as the high-ionisation optical lines. 

\subsubsection{Spectral Evolution}

The optical spectra in this paper span one of the longest periods of any SN.  Figure \ref{fig:spec_evol} highlights only a very slow evolution of the spectrum over three epochs: 736, 3099, and 5105\,d.  Unlike most other SNe, there is no obvious transition to a nebular spectrum even at this extremely late stage. However, the line ratios do exhibit some interesting quantitative changes that are consistent with the shock wave propagating into a medium of decreasing density.

We start with a look at some with the most characteristic SN~IIn lines: the H$\alpha$ to H$\beta$ ratio.  \citet{smith17} showed this ratio was 10.8 at 109\,d and 14.3 at 736\,d. It decreases from  $\sim 24$ to  $\sim 9.3$ to  $\sim 5.4$ at days 905, 3099, and 5105, respectively. We note that the H$\beta$ line is in a region of strong Fe~II lines, which makes the background flux of this line difficult to estimate, especially for the spectrum at 905\,d. The H$\alpha$ to H$\beta$ ratio for this epoch is therefore especially uncertain.  As discussed in Section \ref{sec_lowion}, the general evolution of these line ratios indicates a decreasing optical depth in the Balmer lines.  

\begin{figure}
\centering
\includegraphics[width=3.75in]{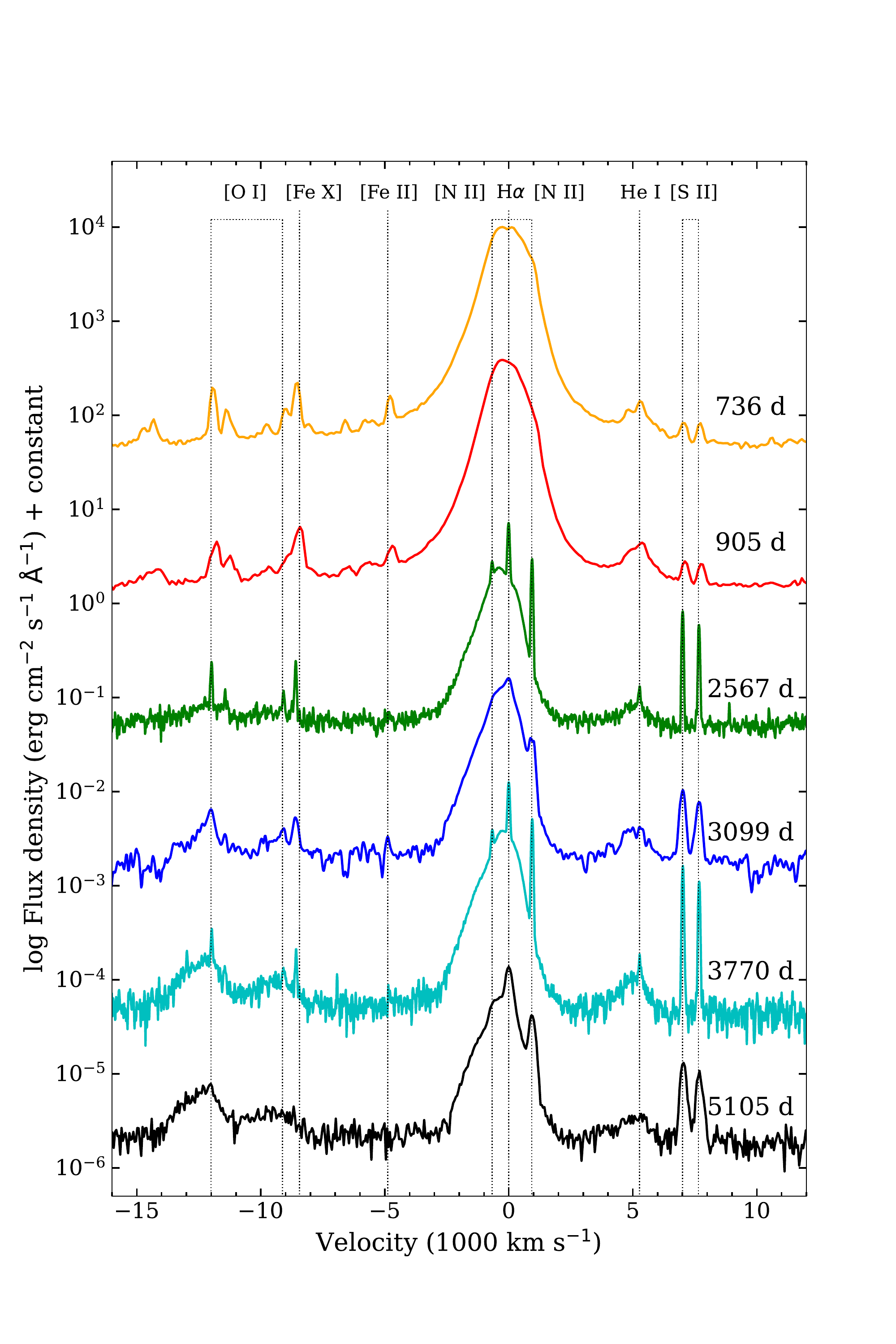}
\caption{Evolution of the H$\alpha$ and [O~I] line profiles from day 736 to day 5105.  The flux of each spectrum is normalised, put on a logarithmic scale, and shifted by a constant. Note the nearly exponential profile on the blue side of the lines at the last four epochs, characteristic of electron scattering. Also note the emergence of the broad [O~I] lines at $\sim 3000$\,d and the fading of the [Fe~X] line in the last spectrum. The smaller FWHM of the narrow lines in the day 2567 and 3770 spectra is caused by the higher spectral resolution in these data. }
\label{fig:Ha_evol}
\end{figure}

The H$\alpha$ blue wing has an exponential profile extending to $\sim 2700$\,\kms (see Figure \ref{fig:Ha_evol}), indicating that electron scattering could still be important \citep{huang18}.  The line has a visible deficit on the red side throughout the whole period, caused by either occultation from an optically thick photosphere or dust. 
The day 736 and 905 day spectra show a faint wing to higher velocity which is gradually fading away. Apart from this, the H$\alpha$ line profile changes little over the period 905 to 5105 days . \cite{stritzinger12} find a drop in the H$\alpha$ flux by factor $\sim 10$ from 900 to 1800 days, and as shown from the r-band photometry in Figure \ref{fig_photometry}, this trend has continued to our last observaitons. 

The He~I $\lambda 7065/\lambda 5876$ ratio decreases from $\sim 0.56$ to $\sim 0.32$ from 905\,d to 5105\,d, again indicating a decrease in optical depth, similar to what we find for the Balmer lines. It is therefore most useful to compare individual line strengths to He~I $\lambda 5876$, which is only moderately affected by optical-depth effects.  The He~I $\lambda 5876$ line has a very similar profile and evolution to H$\alpha$.  Both the [O~I] $\lambda \lambda 6300$, 6364 and the [Ca~II] $\lambda \lambda 7292$, 7324 lines are relatively weak at 905\,d, but increase steadily relative to H$\alpha$ over this period. This trend is again likely a result of decreasing electron density and, in turn, decreasing the effects of collisional de-excitaion. The width of the [O~I] $\lambda 6300$ line is similar to H$\alpha$ at these epochs. The Mg~I] $\lambda 4571$ line, commonly seen in nebular spectra, is not detected above the noise, which is probably a result of the high UV flux above 7.65\,eV, the ionisation threshold of Mg~I (see Figure \ref{fig_stis_spectrum}). 

All of the high-ionisation lines decrease relative to the He~I $\lambda 5876$ line. While the decrease is modest between 905 and 3099\,d, it is steeper between 3099 and 5105\,d (see Fig. \ref{fig:Ha_evol} for the [Fe~X] $\lambda 6374.5$ line) . Typically, the relative fluxes decrease by a factor of $\sim 2$ over this time period.  By contrast, the ratios of the [Fe~VII] lines change little over time.  At 173\,d  $F(\lambda 3586)/F(\lambda 6087) \approx 0.94 \pm 0.32$ and  $F(\lambda 5721)/F(\lambda 6087) \approx 0.58 \pm 0.11$ \citep{smith09ip}, which are similar to the ratios at 905 and 3099\,d. 

In the UV, Figure \ref{fig_stis_spectrum} compares the day 3065 and 4368 spectra.  The day 4368 spectrum exhibits the same lines as the day 3065 spectrum, although the decreasing flux has caused several of the weaker lines to fade below the noise (Figure \ref{fig_stis_spectrum}). The weakest lines include in particular \ion{O}{III}] $\lambda$1664, \ion{N}{III}] $\lambda\lambda$1746.8--1754.0, and \ion{C}{III}] $\lambda\lambda$1906.7, 1908.7, and a meaningful estimate of the CNO abundances can therefore not be made for this later epoch. The \ion{N}{IV}] $\lambda\lambda$1483.3, 1486.5 lines are, however, still detected with a factor of $\sim 3.0$ lower flux compared to day 3065. Also, the \ion{N}{V} $\lambda\lambda 1238.8$, 1242.8 doublet can be seen with a factor of $\sim 6.7$ lower flux. The strongest UV line is still the \ion{Mg}{II} $\lambda\lambda$2795.5, 2802.7 doublet, which has decreased by a smaller factor, $\sim 1.7$.

The lines with the highest ionisation stage (i.e., \ion{N}{V} $\lambda\lambda$1238.8, 1242.8) decrease fastest, followed by \ion{N}{IV}] $\lambda\lambda$1483.3, 1486.5. This trend again illustrates that the general state of ionisation in the lower density CSM has decreased considerably between the two epochs. This is consistent with the steadily decreasing X-ray luminosity (Figure \ref{fig_photometry}).  Many of the higher ionisation lines also disappeared or weakened around this same epoch in SN 1988Z \citep{smith17}.
 
\subsection{CNO Processing Abundances}
\label{sec:cno}

\begin{table}
\centering
\caption{ Ionic abundance ratios \label{tab_abundances}}
\begin{tabular}{ c c c c}
\hline
\hline
\ion{N}{III}/\ion{C}{III}	&	\ion{N}{IV}/\ion{C}{IV}	&	\ion{N}{III}/\ion{O}{III}	&	\ion{N}{IV}/\ion{O}{III}\\
\hline
9.5--12.8 $\pm$ 2.3 & 15.7--17.1 $\pm$ 2.0 & $>$ 3.3--6.0	& 1.2--2.0\\
\end{tabular}
\end{table}

The relative abundances of ions listed in Table \ref{tab_uvlineflux} depend on the temperatures and densities of the CSM.  These lines have similar excitation energies, however, so the temperature sensitivity is relatively weak.  If the density is less than the critical densities of these semiforbidden lines ($\sim10^9$\,cm$^{-3}$), then the density sensitivity is also weak.

On day $\sim 100$, \citet{smith09ip} use the critical densities of the high-ionisation optical coronal lines to set an upper limit on the density of the pre-shocked CSM at $<10^7$--$10^8$\,cm$^{-3}$. The absence of the [O~II] $\lambda\lambda$3726, 3729 doublet before day 173 sets a lower limit on the density of $\gtrsim 3 \times 10^5$\,cm$^{-3}$. However, the  [\ion{S}{II}] $\lambda\lambda$6717, 6731 doublet ratio on day 173 yields a density of only $\sim10^2$\,cm$^{-3}$.  \citet{smith09ip} suggest that this inconsistency may indicate an inhomogeneous CSM.  These measurements correspond to days 109--173, so we may infer the densities by day 3000 could be lower by a factor of $\sim(3000/100)^2 \approx 10^2$ for a steady wind. However, a steady wind may not apply for a time-limited eruption and clumpiness further complicates this.

Using the above temperature and density information, we calculate the elemental abundances using the ionic flux ratios from Table \ref{tab_uvlineflux} and atomic data from the CHIANTI compilation \citep{landi12}.  The values given correspond to ratios using temperatures in the range (1.0--3.0) $\times 10^4$\,K and densities up to $10^6$\,cm$^{-3}$.  To calculate the elemental abundances, however, requires assumptions about the ionisation structure of the CSM.  X-ray photoionised models characterised by a 10\,keV bremsstrahlung spectrum reveal that the \ion{C}{III} and \ion{N}{III} zones, as well as the \ion{C}{IV}, \ion{N}{IV}, and \ion{O}{III} zones, nearly coincide \citep{kallman82}.  SN 2005ip has a somewhat different ionising spectrum and a lower CSM density, so this model is not necessarily true. However, the presence of numerous high-ionisation lines, like [\ion{Ar}{X}] $\lambda$5536, as well as the direct observations of a high X-ray luminosity, argue for X-ray ionisation to dominate, although the exact spectrum is uncertain.  Because more detailed photoionisation models do not exist, we assume here a similar ionisation structure as described by \citet{kallman82}.

Table \ref{tab_abundances} lists the abundance ratios of \ion{C}{III-IV}, \ion{N}{III-IV}, and \ion{O}{III}: N/C = 9.5--12.8 $\pm$ 2.3 and N/O $>$ 1.2--2.0.  Since the \ion{C}{IV} line is severely suppressed by ISM absorption from the host galaxy, we did not include the \ion{N}{IV}/\ion{C}{IV} ratio in the analysis.  Owing to the closer correspondence of the \ion{N}{IV} and \ion{O}{III} zones in the models, we prefer to use these ions to calculate the N/O ratio.  Independent of uncertainties in densities, temperatures, and ionisation structure, these results show that the CSM of SN 2005ip is strongly N enriched.  Compared to solar \citep{Asplund09} the above ratios are enhanced by a factors of 38--51 for N/C and $> 8.7$--14 for N/O, consistent with products from CNO burning. 

\subsection{Grain Heating and Dust Mass}

The {\it Spitzer} data can provide important constraints on the dust heating and total dust mass. The assumed physical scenario is that a dense, external dust shell was formed during a pre-SN eruption and is now heated by internal optical, UV, and X-ray emission generated by shock interaction \citep[Section 3.1,][]{fox11}.  We use the three-dimensional radiative transfer code Monte Carlo Simulations of Ionized Nebulae \citep[MOCCASIN][and references within]{ercolano05} to analyze data during the {\it Spitzer} epochs closest in time to the two epochs of {\it HST}/STIS UV spectroscopy (e.g., 2015 and 2018). MOCASSIN accepts multiple parameters defined by a user and executes the desired scenario on a Cartesian grid. Both epochs were initially modeled with a simple blackbody and then again with a modified blackbody.  We assume a spherical shell with 100\% amorphous carbon dust distributed uniformly and a standard \citet[][MRN]{mathis77} grain size distribution of $a^{-3.5}$ between 0.005 and 0.05\,$\mu$m.  The model accounts only for the observed dust, which is most likely the hottest dust at the inner radius of the external shell of material.  A much higher mass (10--100 times more) of colder dust likely exists \citep[e.g.,][]{matsuura17}.

The 2018 epoch was modeled first since it has accompanying ground-based optical photometry from the LBT and NIR photometry from UKIRT.  The best-fit model suggests a shell with $R_{\rm in} = 9.0 \times 10^{15}$\,cm and $R_{\rm out} = 6.0 \times 10^{16}$\,cm. The dust temperature ranges from 700\,K to 600\,K at the inner and outer edges, respectively.  The implied dust mass is $1.59 \times 10^{-3}$\,M$_\odot$ with an equivalent optical depth of 2.46. The addition of the UV spectrum resulted in a minimal increase in the dust temperature ($\Delta T \approx 10$\,K) and infrared brightness ($\Delta  F_\nu \approx 0.01$\,mJy).

For the 2015 spectrum, we used the same smooth shell model derived for the 2018 fit and an input luminosity of $1.35 \times 10^{41}$\,erg\,s$^{-1}$ ($\sim 3.51 \times 10^7$\,L$_\odot$).  The temperature of the dust was found to range from about 720\,K to 530\,K from the inner to outer radius of the shell, respectively. The implied dust mass and equivalent optical depth were also the same as in the 2018 model. The addition of the UV spectrum to the model resulted in slightly higher dust temperatures ($\Delta T \approx 10$ to 40\,K, by grain size) and infrared brightness ($\Delta F_\nu \approx 0.1$\,mJy).

\subsection{The Dust Powering Mechanism}

Following a nearly 2000\,d plateau, the MIR evolution qualitatively follows a similar slow decline as the optical of $\sim 1$\,mag\,(1000\, d)$^{-1}$, although there may be an indication of a slower decline at the latest epochs.  The MIR {\it Spitzer} photometry from two earlier epochs (days 948 and 2057) are already shown to be consistent with a large, pre-existing dust shell that is continuously heated by visible and X-ray radiation generated by ongoing CSM interaction \citep{fox10,fox11}.  Assuming a spherically symmetric dust shell located at the blackbody radius, the MIR would require a visible/X-ray heating luminosity from CSM interaction of $\sim10^8$\,\lsolar\ at these epochs (see Figure 10b of \citealt{fox11}).  This luminosity is consistent with the $r$-band photometry on day 948 given in Figure 10 of \citet{smith09ip} or from Figure 9 of \citet{stritzinger12}, based on the full optical through IR SED.

Another consistency check requires that the optical depth, $\tau$, for such a system be $> 1$.  Following Equation 23 from \citet{dwek17}, the dust's optical depth at wavelength $\lambda$ can be written as
\begin{equation}
\tau(\lambda) = Z_{\rm dH} m_H \kappa_{\rm ext}(\lambda) N_{\rm H},
\end{equation}
where $Z_{\rm dH}$ is the dust-to-H mass ratio, $m_H$ is the hydrogen atomic mass, $\kappa_{\rm ext}$ is the extinction coefficient, and $N_{\rm H}$ is the column depth.  The X-ray observations in both this paper and \citet{katsuda14} show a column depth, $N_{\rm H}$, that decreases from $\sim 5 \times 10^{22}$\,cm$^{-2}$ to $0.1 \times 10^{22}$\,cm$^{-2}$ over time.  From Figure 4 of \cite{fox10} we assume $10^4< \kappa_{\rm ext}(V) < 10^6$\,cm$^{-2}$\,g, which represents a typical value for a variety of grain sizes $<1.0$\,\micron\ and compositions. For these values, we derive $\tau(V) \approx 1.7 \times 10^2\,Z_{\rm dH}$.  For typical gas-to-dust mass ratios of $\sim 100$ ($Z_{\rm dH} \approx 0.01$), this implies an optical depth $\tau > 1$ throughout the latest epochs for most values of $\kappa$, typically corresponding to smaller grains.  While these are order-of-magnitude approximations, a value $\tau > 1$ for a spherical shell would impact the measured extinction and total radiated energy.  The fact that both the measured extinction is low \citep{stritzinger12} and the optical to mid-IR hovers around $\sim 1$ suggests that the geometry is likely more complicated than a spherical shell, perhaps clumpy or an anisotropic shell or disk \citep{smith09ip,fox09,fox10,stritzinger12,smith17}.

\section{Summary}
\label{sec:summary}

We have presented very late-time observations of SN 2005ip, including {\it Chandra}/ACIS X-ray spectra, {\it HST}/STIS UV spectra, {\it HST}/WFC3 optical photometry, ground-based optical and NIR photometry, and {\it Spitzer}/IRAC photometry.  The MIR evolution is consistent with the previously proposed pre-existing dust shell that is radiatively heated by ongoing CSM interaction.  There may be some indication of a relatively slower MIR decline in the latest epochs, which could suggest a slowly decaying thermal echo.  The total energy radiated by the shock so far is in excess of $10^{50}$\,erg.   The large energy release indicates an efficient conversion of  kinetic energy to radiation.  The progenitor mass-loss rate we find is $\ga 1 \times 10^{-2}\,{\rm  M_{\odot}\,yr}^{-1}$, and the total mass lost is $\ga 1\,{\rm M_\odot}$ but can be considerably larger, depending on the exact efficiency for the conversion of shock energy to radiation.  This explosion is thought to arise from a massive, luminous red supergiant, like VY~CMa \citep{smith09rsg}, or an LBV progenitor in order to account for the huge mass-loss rate \citep{fox09,smith09ip,stritzinger12}.  The optical spectra show strong effects of Ly$\alpha$ fluorescence and a decreasing optical depth in the lines.   The UV spectra show that the CSM of SN 2005ip is strongly N enriched, consistent with products from CNO burning. 

After more than 5\,yr of a relatively flat plateau, the light curve has begun to fade in all bands.  This result indicates that the shock may finally be reaching the outer extent of the dense CSM shell around SN~2005ip.  The final optical and MIR photometry, however, leaves some ambiguity that the declining CSM interaction could still be continuing at a lesser strength, and only time will tell if the demise is permanent.\\

\noindent{\bf Acknowledgements}

We thank the anonymous referee for suggestions that improved this paper. Some of the data presented herein were obtained at the W. M. Keck Observatory, which is operated as a scientific partnership among the California Institute of Technology, the University of California, and the National Aeronautics and Space Administration (NASA); the Observatory was made possible by the generous financial support of the W. M. Keck Foundation. The authors wish to recognise and acknowledge the very significant cultural role and reverence that the summit of Maunakea has always had within the indigenous Hawaiian community; we are most fortunate to have the opportunity to conduct observations from this mountain. We thank WeiKang Zheng for helping obtain Keck optical photometry and spectroscopy.
When the data reported here were acquired, UKIRT was supported by NASA and operated under an agreement among the University of Hawaii, the University of Arizona, and Lockheed Martin Advanced Technology Center; operations were enabled through the cooperation of the East Asian Observatory. The LBT is an international collaboration among institutions in the United States, Italy, and Germany. The LBT Corporation partners are The University of Arizona on behalf of the Arizona university system; Istituto Nazionale di Astrofisica, Italy; LBT Beteiligungsgesellschaft, Germany, representing the Max Planck Society, the Astrophysical Institute Potsdam, and Heidelberg University; The Ohio State University; The Research Corporation, on behalf of The University of Notre Dame, University of Minnesota and University of Virginia. Observations reported here were obtained at the MMT Observatory, a joint facility of the University of Arizona and the Smithsonian Institution. This publication makes use of data products from the Two Micron All Sky Survey, which is a joint project of the University of Massachusetts and the Infrared Processing and Analysis Center/California Institute of Technology, funded by NASA and the U.S. National Science Foundation (NSF).

This work is based in part on observations obtained with the {\it Spitzer Space Telescope}, which is operated by the Jet Propulsion Laboratory, California Institute of Technology, under a contract with NASA. Financial support for this work was provided by NASA through grants GO-10877, GO-13287,  GO-14598, GO-14688, and GO-15166 from the Space Telescope Science Institute (STScI), which is operated by the Associated Universities for Research in Astronomy, Inc. (AURA), under NASA contract NAS 5-26555. A.V.F.'s supernova group has also been supported by NASA/{\it Chandra} grant GO7-18067X, the Christopher R. Redlich Fund, the TABASGO Foundation, NSF grant AST-1211916, and the Miller Institute for Basic Research in Science (U.C. Berkeley). C.F. acknowledges support from the Swedish Research Council and Swedish National Space Board. M.D.S. is supported by generous grants (13261 and 28021) from VILLUM FONDEN and by a project grant (8021-00170B) from the Independent Research Fund Denmark. T.S. is supported by the GINOP-2-3-2-15-2016-00033 project (``Transient Astrophysical Objects'') of the National Research, Development and Innovation Office (NKFIH), Hungary, funded by the European Union.\\

\noindent{\bf Data availability}\\

The data underlying this article will be shared on reasonable request to the corresponding author.\\

\bibliographystyle{mnras}
\bibliography{references2}

\end{document}